\documentclass[english,journal,12pt,draftclsnofoot,onecolumn]{IEEEtran}
\usepackage[T1]{fontenc}
\usepackage[latin9]{inputenc}
\usepackage[letterpaper]{geometry}
\usepackage{color}
\usepackage{array}
\usepackage{float}
\usepackage{multirow}
\usepackage{amsmath}
\usepackage{amsthm}
\usepackage{amssymb}
\usepackage{stmaryrd}
\usepackage{graphicx}
\usepackage{esint}

\makeatletter

\providecommand{\tabularnewline}{\\}
\floatstyle{ruled}
\newfloat{algorithm}{tbp}{loa}
\providecommand{\algorithmname}{Algorithm}
\floatname{algorithm}{\protect\algorithmname}

\theoremstyle{plain}
\newtheorem{thm}{\protect\theoremname}
\theoremstyle{remark}
\newtheorem{rem}[thm]{\protect\remarkname}

\usepackage[caption=false,font=footnotesize]{subfig}
\usepackage{algorithm}
\usepackage{algorithmicx}
\usepackage{algpseudocode} 
\@ifundefined{showcaptionsetup}{}{%
 \PassOptionsToPackage{caption=false}{subfig}}
\usepackage{subfig}
\makeatother

\usepackage{babel}
\providecommand{\remarkname}{Remark}
\providecommand{\theoremname}{Theorem}

\begin{document}
\title{Hybrid Driven Learning for Channel Estimation in Intelligent Reflecting
Surface Aided Millimeter Wave Communications}
\author{Shuntian~Zheng,~Sheng~Wu,~\IEEEmembership{Member,~IEEE,}~Chunxiao~Jiang,~\IEEEmembership{Senior~Member,~IEEE,}~\\
 Wei~Zhang,~\IEEEmembership{Fellow,~IEEE,}~and~Xiaojun~Jing,~\IEEEmembership{Member,~IEEE}\thanks{Shuntian~Zheng, Sheng~Wu and Xiaojun~Jing are with the School of
Information and Communication Engineering, Beijing University of Posts
and Telecommunications, Beijing 100876, China (e-mail: %
\mbox{%
\{shuntianzh, thuraya, jxiaojun\}%
}@bupt.edu.cn).

Chunxiao~Jiang is with the Tsinghua Space Center, Tsinghua University,
Beijing 100084, China, and also with the Beijing National Research
Center for Information Science and Technology, Tsinghua University,
Beijing 100084, China (e-mail: %
\mbox{%
jchx%
}@tsinghua.edu.cn).

Wei~Zhang is with the School of Electrical Engineering and Telecommunications,
University of New South Wales, Sydney, NSW 2052, Australia (e-mail:
\mbox{%
w.zhang%
}@unsw.edu.au).}}

\maketitle
\vspace{-0.8cm}
\begin{abstract}
Intelligent reflecting surfaces (IRS) have been proposed in millimeter
wave (mmWave) and terahertz (THz) systems to achieve both coverage
and capacity enhancement, where the design of hybrid precoders, combiners,
and the IRS typically relies on channel state information. In this
paper, we address the problem of uplink wideband channel estimation
for IRS aided multiuser multiple-input single-output (MISO) systems
with hybrid architectures. Combining the structure of model driven
and data driven deep learning approaches, a hybrid driven learning
architecture is devised for joint estimation and learning the properties
of the channels. For a passive IRS aided system, we propose a residual
learned approximate message passing as a model driven network. A denoising
and attention network in the data driven network is used to jointly
learn spatial and frequency features. Furthermore, we design a flexible
hybrid driven network in a hybrid passive and active IRS aided system.
Specifically, the depthwise separable convolution is applied to the
data driven network, leading to less network complexity and fewer
parameters at the IRS side. Numerical results indicate that in both
systems, the proposed hybrid driven channel estimation methods significantly
outperform existing deep learning-based schemes and effectively reduce
the pilot overhead by about 60\% in IRS aided systems. 
\end{abstract}

\begin{IEEEkeywords}
Hybrid driven, channel estimation, Intelligent reflecting surfaces,
deep learning.
\end{IEEEkeywords}

\IEEEpeerreviewmaketitle{}

\section{Introduction}

Recently, intelligent reflecting surfaces (IRS) have gained significant
attention. It has been recognized as a potential technology for beyond
5G/6G wireless communications \cite{IRS_Survey}. In contrast to traditional
relaying systems, an IRS consists of a large number of nearly-passive
reflectors and an IRS controller, where the IRS controller independently
controls the amplitude and phase shift of each reflector to adapt
the wireless propagation environment between the base station (BS)
and the user equipment (UE) \cite{IRS_access}. The throughput enhancement
becomes significant via virtual line-of-sight (LoS) links, and electromagnetic
materials allow IRS to be flexibly integrated into existing networks
(e.g., cellular networks) \cite{IRS_Survey}. Given these attractive
characteristics, IRS has received considerable research interests
in both academia and industry, especially in terms of energy efficiency
spectral efficiency maximization, received signal-to-noise ratio (SNR)
enhancement \cite{TWC19Rui} and physical layer security \cite{pls_irs},
etc.

\subsection{Motivation and Related Work}

Numerous works addressed the optimization of the reflection coefficient
vector at the IRS and the hybrid precoding matrix at the BS \cite{TWC19Rui,MA_IRS_THz}.
However, to fully utilize the potential of the IRS in communication
systems, the BS and IRS controller require to design the precoding
matrix and reflection parameters, which depend on accurate channel
state information (CSI). Furthermore, the large number of antenna
arrays and passive reflecting elements increases the estimation overhead.
These challenges motivate researchers to investigate the properties
of IRS aided channels for accurate channel estimation. Some conventional
estimators (e.g., a least squares estimator was proposed by Wang \emph{et
al}. \cite{wang2020compressed} and a linear minimum mean square error
estimator was studied by Liu \emph{et al}. \cite{CDRN}) require a
large number of pilots. Exploiting the common spatial sparsity, Chen
\emph{et al}. \cite{irs_struct} formulated the BS-IRS-UE cascaded
estimation as a sparse recovery problem and leveraged compressive
sensing (CS) methods. Beyond the conventional estimation algorithms,
there are various estimation frameworks to assist in solving the channel
estimation problem for multi-user IRS cascade channels \cite{Hu_twoscale}.
Specifically, the high-dimensional cascaded channel is split into
BS-IRS and IRS-UE channels, which are estimated respectively. However,
the BS-IRS channel may be non-static due to the high mobility of BS
or IRS (e.g., IRS enhanced unmanned aerial vehicles aided networks). 

In recent years, the massive hybrid massive multiple-input multiple-output
(MIMO) architecture with a much smaller number of radio frequency
(RF) chains has been widely considered for massive MIMO systems \cite{OMP_CE_17,SOMP18}.
Up to now, the research on frequency-selective channel estimation
framework for IRS aided communication systems with a hybrid architecture
is limited. The problem of channel estimation in the IRS aided hybrid
massive MIMO orthogonal frequency division multiplexing (OFDM) system
is even more problematic. In practice, the observations are limited
when the number of RF chains is much smaller than the number of antennas.
Zhang \emph{et al}. \cite{irs-hybrid-ce} studied a linear digital
estimator for the IRS aided narrowband channel recovery problem. Nevertheless,
the channel estimation accuracy depends on the channel sparsity, which
is usually unknown. 

Alternatively, deep learning approaches have proven their effectiveness
in wireless communication, such as signal detection \cite{DNN-CE}
and passive beamforming design \cite{hybrid_IRS}. Some learning networks
based on data driven \cite{CDRN,GAN_CE,att_ce} and model driven algorithms
\cite{LAMP,LDAMP} have been applied in channel estimation or signal
detection in the mmWave band. In specific, data driven approaches
extract features of massive training data to improve the performance.
Exploiting massive training data and neural networks, including convolutional
neural networks (CNNs) or fully-connected neural networks, the data
driven approaches assist or design the estimation models \cite{CDRN,GAN_CE}.
Inspired by traditional estimation models and the powerful learning
capabilities of neural networks, several model driven approaches provide
appealing networks to extend estimation models. A few recent estimators
used a learned approximate message passing (LAMP) network by unfolding
the AMP algorithm \cite{wu_twc} to multiple neural layers \cite{LAMP,LDAMP,GMAMP,MMVLAMP}.
In summary, data driven learning networks are regarded as black boxes,
which directly map or enhance the related information to CSI. However,
model driven learning networks learn parameters from the traditional
iterative structures, which neglect some specific properties of the
channel. These works motivate us to exploit both model driven and
data driven structures and learn the characteristics of the high-dimensional
channels.

\subsection{Our Contributions}

In this paper, we propose and study novel hybrid driven channel estimation
algorithms for frequency-selective channels in IRS aided systems.
We consider the uplink estimation for passive and hybrid IRS aided
multiple-input single-output (MISO) communication systems and present
two hybrid driven estimation algorithms for both scenarios. In contrast
to existing learning estimation approaches, e.g., \cite{CDRN,GAN_CE,att_ce,LAMP}
and \cite{MMVLAMP}, the developed methods reap the advantages of
both data driven and model driven methods to estimate wideband mmWave
channel. Although the networks in this paper are deployed for IRS
aided systems, the proposed methods can also be applied to massive
MIMO systems. The contributions are summarized as follows:
\begin{itemize}
\item We model passive and hybrid IRS aided wideband hybrid MISO-OFDM systems
as two sparse recovery problems. In the CS formulation, we apply redundant
dictionary matrices with higher angular resolution. With the aid of
the data driven network, model driven network and residual learning
mechanism, the hybrid driven network architecture is proposed to further
improve the estimation performance and reduce the pilot overhead.
\item For the cascaded channel estimation in a passive IRS aided system,
we develop a denoising and attention mechanism assisted residual learned
approximate message passing (DA-RLAMP) network. Specifically, the
denoising network extracts the noise from the observations, while
the attention network consists of a frequency attention network and
a spatial network to learn and enhance structural characteristics
in frequency and spatial, respectively. 
\item For channel estimation in a hybrid passive and active IRS aided system,
we extend the above framework by trading off the estimation error
and hardware requirement at the IRS side. Specifically, the depthwise
separable convolution is developed to effectively integrate the attention
mechanism into the mobile DA-RLAMP (MDA-RLAMP) network, which reconstructs
the UEs-IRS channel with low complexity and expands the application
range of the proposed approach.
\item To reliably reconstruct the complex-valued channels, two complex-valued
networks are developed in the proposed approaches. We derive an explicit
expression of the proposed algorithms to explain the mechanism of
the proposed network theoretically. Extensive simulations have been
conducted to verify that the proposed algorithms outperform the traditional
estimation methods, data driven networks and model driven networks. 
\end{itemize}
The rest of the paper is organized as follows. Section II presents
the passive and hybrid IRS aided multi-user MISO (MU-MISO) system
model and the high-frequency channel model. Section III presents the
high-frequency channel models and formulates the channel estimation
problems with redundant dictionary matrices. Section IV develops two
hybrid driven channel estimation frameworks for both scenarios. Numerical
results are demonstrated in Section V. Eventually, the work of this
paper is concluded in Section VI.

Notations: We use letters (e.g., $x$), bold lowercase letters (e.g.,
$\boldsymbol{x}$) and bold uppercase letters (e.g., $\boldsymbol{X}$)
to indicate scalars, vectors and matrices, respectively. The superscripts
$\left(\cdot\right)^{\mathsf{T}}$, $\left(\cdot\right)^{\mathsf{*}}$and
$\left(\cdot\right)^{\mathsf{H}}$ are transpose operation, complex
conjugate operation and Hermitian transpose operation, respectively.\emph{
}$\mathcal{\mathsf{diag}}\left\{ \boldsymbol{x}\right\} $ represents
a square diagonal matrix whose main diagonal elements are given by
$\boldsymbol{x}$; $\left\Vert \cdot\right\Vert _{2}$ and $\left\Vert \cdot\right\Vert _{\mathsf{F}}$
denote the $\ell_{2}$ and is the Frobenius norms, respectively. $\mathsf{vec}(\boldsymbol{X})$
and $\mathsf{vec2mat}\left(\boldsymbol{x},\left[m\times n\right]\right)$
correspond to transforming $\boldsymbol{X}$ into a vector and transforming
$\boldsymbol{x}$ into a matrix for a defined size $m\times n$. $\boldsymbol{X}\otimes\boldsymbol{Y}$
and $\boldsymbol{X}\odot\boldsymbol{Y}$ denote Kronecker and Khatri-Rao
products of $\boldsymbol{X}$ and $\boldsymbol{Y}$. |�| represents
determinant or absolute value depending on context. $\mathcal{N}_{\mathbb{C}}\left(\mu,\sigma^{2}\right)$
denotes the complex Gaussian random variable with mean $\mu$ and
variance $\sigma^{2}$. $\mathsf{Re}\left(\boldsymbol{x}\right)$
and $\mathsf{Im}\left(\boldsymbol{x}\right)$ denote  the real and
imaginary parts of $\boldsymbol{x}$.

\section{IRS Aided MISO System \label{sec:System-Model}}

In this section, we first present the IRS aided hybrid MISO system.
Then, we introduce the estimation frameworks for passive and hybrid
IRS, respectively.

\subsection{Passive IRS Aided MU-MISO System \label{subsec:IRS_system}}

\begin{figure*}
\vspace{-0.6cm}

\centering\subfloat[]{\centering\includegraphics[height=1.5in]{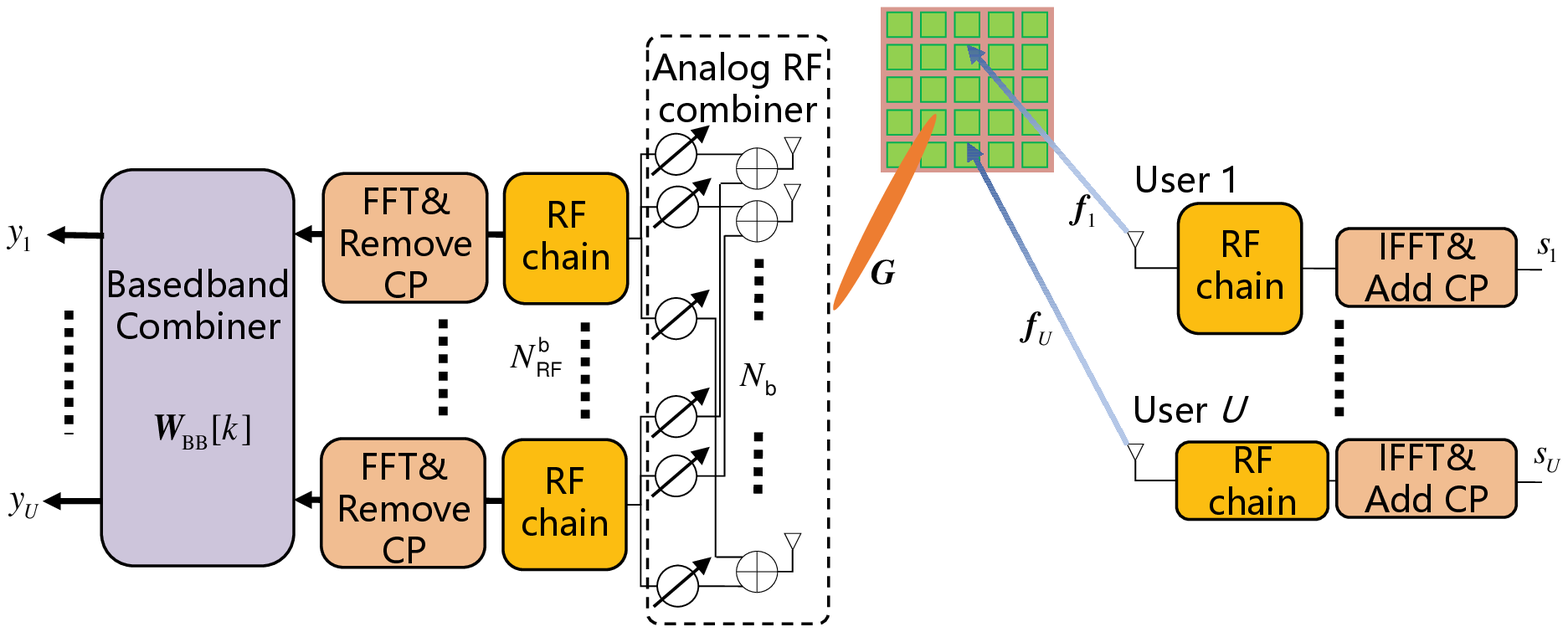}}
\subfloat[]{\centering\includegraphics[height=1.4in]{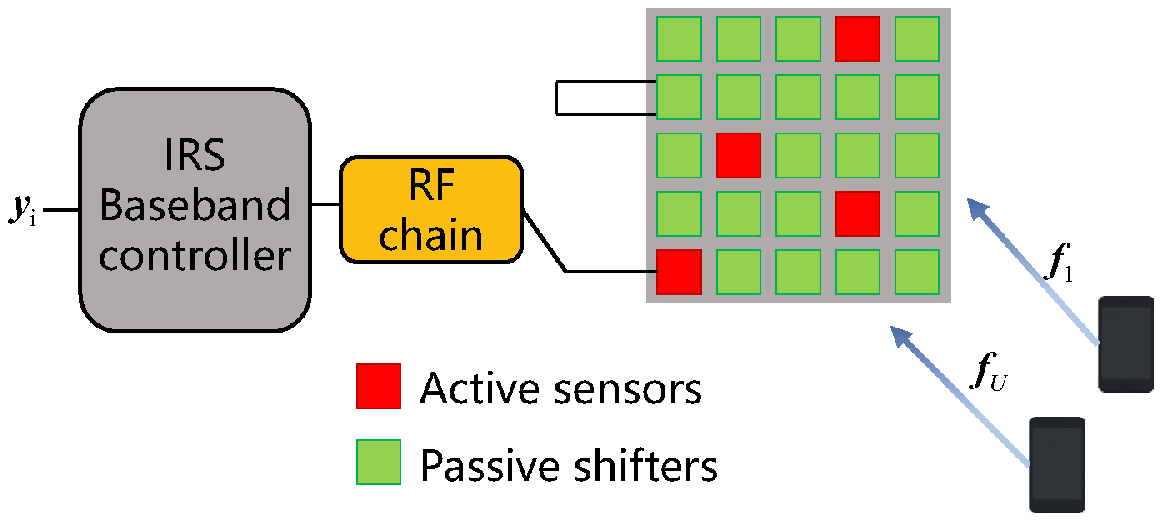}}

\caption{IRS aided communication models. (a) The Passive IRS aided hybrid architecture
communication model. (b) The hybrid passive and active IRS aided communication
model.}

\vspace{-0.8cm}

\label{fig:IRS system}
\end{figure*}

Consider the IRS aided MU-MISO scenario as presented in Fig. \ref{fig:IRS system}
(a). A hybrid precoding structure has been widely considered at the
BS side, which generates directional beams with high array gain in
Massive MIMO systems \cite{SOMP18}. The phase shift of each reflective
element on the IRS is configurable via an intelligent controller,
which obtains the CSI of all UEs from the BS \cite{TWC19Rui}. Then,
the hybrid precoding at the BS and phase shift at IRS are jointly
optimized during the downlink transmission. Therefore, we focus on
a time division duplexing system and formulate the estimation problem
at the BS side.\emph{ }The uplink OFDM signals employ total $K$ $(k=1,2,\cdots K)$
subcarriers to send pilot symbols from $U$ UEs with a single antenna
to the BS with $N_{\mathsf{b}}$ antennas. The BS is equipped with
$N_{\mathsf{RF}}^{\mathsf{b}}=U$ radio frequency (RF) chains and
each user has a single RF chain\emph{}\footnote{Note that the estimation problem is also applicable in the case of
the IRS aided MU-MISO-OFDM system relying on the lens antenna array.}. 
The passive IRS is installed with $N_{\mathsf{i}}$ passive phase
shifters to provide a LoS link between the BS and the UEs. Each shifter
in the IRS combines and reflects all the signals by dynamically adjusting
the phase. During the $t^{\mathsf{th}}(t=1,2\ldots,T)$ pilot symbol,
each shifter applies a new phase instantaneously without loss of generality.
Thus the phase shift matrix of each IRS element in $t^{\mathsf{th}}$
symbol is denoted by
\begin{equation}
\boldsymbol{r}^{\left(t\right)}=\left[\beta^{\left(t\right)}e^{j\varphi_{1}^{(t)}},\beta^{\left(t\right)}e^{j\varphi_{2}^{(t)}},\ldots,\beta^{\left(t\right)}e^{j\varphi_{N_{\mathsf{i}}}^{(t)}}\right]^{\mathsf{T}}\in\mathbb{C}^{N_{\mathsf{i}}},
\end{equation}
where $\beta^{\left(t\right)}\in\left[0,1\right]$ denotes the amplitude
of the $n_{\mathsf{i}}^{\mathsf{th}}$ $(n_{\mathsf{i}}=1,2\ldots,N_{\mathsf{i}})$
IRS shifter, and $\varphi_{n_{\mathsf{i}}}^{(t)}\in\left[0,2\pi\right]$
is the corresponding phase.\emph{ }The BS first combines the received
signals with an analog combiner $\boldsymbol{W}_{\mathsf{RF}}^{(t)}\in\mathbb{C}^{N_{\mathsf{b}}\times N_{\mathsf{RF}}^{\mathsf{b}}}$
where $N_{\mathsf{s}}=N_{\mathsf{RF}}^{\mathsf{b}}$ denotes data
streams from BS to UEs. Then, the signals are transformed into the
frequency-domain using $N_{\mathsf{RF}}^{\mathsf{b}}$ parallel $K$-point
fast fourier transform and combined with digital baseband combiner
$\boldsymbol{W}_{\mathsf{BB}}^{(t)}[k]\in\mathbb{C}^{N_{\mathsf{RF}}^{\mathsf{b}}\times N_{\mathsf{s}}}$.
During the $t^{\mathsf{th}}$ transmission symbol, the $k^{\mathsf{th}}$
subcarrier received signals $\boldsymbol{y}^{\left(t\right)}[k]\in\mathbb{C}^{N_{\mathsf{s}}\times1}$
is modeled by
\begin{equation}
\boldsymbol{y}^{\left(t\right)}[k]=\boldsymbol{W}_{\mathsf{b}}^{\left(t\right)}[k]^{\mathsf{H}}\sum_{u=1}^{U}\boldsymbol{H}_{u}^{(t)}[k]s_{u}^{\left(t\right)}[k]+\boldsymbol{W}_{\mathsf{b}}^{\left(t\right)}[k]^{\mathsf{H}}\boldsymbol{n}^{(t)}[k],\label{eq:Y_U}
\end{equation}
where $\boldsymbol{W}_{\mathsf{b}}^{\left(t\right)}[k]=\boldsymbol{W}_{\mathsf{RF}}^{\left(t\right)}[k]\boldsymbol{W}_{\mathsf{BB}}^{(t)}$
denotes the hybrid combiner and $\boldsymbol{n}^{(t)}[k]\sim\mathcal{N}_{\mathbb{C}}\left(0,\sigma^{2}\mathbf{I}\right)$
is the additive white Gaussian noise with covariance $\sigma^{2}$
at the BS side; $s_{u}^{\left(t\right)}[k]$ denotes the transmitted
pilot symbol subject to $\mathbb{E}\left[s_{u}^{\left(t\right)}[k]s_{u}^{\left(t\right)}[k]^{\mathsf{H}}\right]=1$.
For simplicity, we assume that the direct link is blocked due to adverse
propagation conditions \cite{wang2020compressed}. The $\boldsymbol{H}_{u}^{\left(t\right)}[k]$
represents the effective frequency-domain channel response from the
$u^{\mathsf{th}}$ user to the BS at the $k^{\mathsf{th}}$ subcarrier,
which is equivalently written by 
\begin{equation}
\boldsymbol{H}_{u}^{\left(t\right)}[k]=\boldsymbol{G}[k]\left(\mathcal{\mathsf{diag}}\left(\boldsymbol{r}^{\left(t\right)}\right)\boldsymbol{f}_{u}[k]\right),\label{eq:cascaded channel}
\end{equation}
where $\boldsymbol{G}[k]\in\mathbb{C}^{N_{\mathsf{b}}\times N_{\mathsf{i}}}$
and $\boldsymbol{f}_{u}[k]\in\mathbb{C}^{N_{\mathsf{i}}\times1}$
stand for the reflected channel between BS and IRS and the direct
channel between IRS and $u^{\mathsf{th}}$ user, respectively. Moreover,
the UEs-IRS-BS cascaded channel can be modeled as 
\begin{equation}
\boldsymbol{H}_{\mathsf{c},u}[k]=\boldsymbol{G}[k]\mathcal{\mathsf{diag}}\left(\boldsymbol{f}_{u}[k]\right).\label{eq:cascaded channel_1}
\end{equation}

\subsection{Hybrid IRS Aided MU-MISO System \label{subsec:Hybrid IRS}}

The pilot overhead of cascaded UEs-IRS-BS estimation is proportional
to the product of the number of IRS shifters and BS antennas, which
yields extremely huge estimation time at the BS side. On the other
hand, the performance gain of passive IRS is fundamentally limited
by the severe multiplicative path loss of the cascaded channel \cite{Hu_twoscale}.
To overcome the above drawbacks, a hybrid IRS architecture is developed
in \cite{Tensor_IRS} and \cite{hybrid_IRS}, as shown in Fig. \ref{fig:IRS system}
(b), which is composed of passive shifters and active sensors. Furthermore,
$N_{\mathsf{i}}^{a}$ active sensors consist of both estimation and
reflection phases. Considering the cost of high power consumption
and hardware constraints, the active sensors connect to the baseband
unit via an RF chain and directly sense the dynamic UEs-IRS channels
by observing the received pilots in the estimation phase \cite{hybrid_IRS}.
During the data transmission phase, the passive shifters and active
sensors reflect the signals wave simultaneously. In a hybrid IRS system,
we focus on the mobile UEs-IRS channel, where the received signals
at the $t_{\mathsf{i}}^{\mathsf{th}}(t=1,\ldots,T_{\mathsf{i}})$
training symbol can be expressed as \cite{Lin_TWC18}
\begin{equation}
\boldsymbol{y}_{\mathsf{i}}^{\left(t_{\mathsf{i}}\right)}[k]=\boldsymbol{w}_{\mathsf{i}}^{\left(t_{\mathsf{i}}\right)}[k]^{\mathsf{H}}\sum_{u=1}^{U}\boldsymbol{f}_{u}[k]s_{u}^{\left(t_{\mathsf{i}}\right)}[k]+\boldsymbol{W}_{\mathsf{i}}^{\left(t_{\mathsf{i}}\right)}[k]^{\mathsf{H}}\boldsymbol{n}^{\left(t_{i}\right)}[k],\label{eq:hybrid IRS Y}
\end{equation}
where $\boldsymbol{w}_{\mathsf{i}}\in\mathbb{C}^{N_{\mathsf{i}}\times1}$
stacks a one-hot vector, with 0 indicating a passive shifter and 1
representing an active sensor.

\section{Channel Model and Problem Formulation \label{sec:System-Model-1}}

In this section, the channel model of the UEs-IRS and IRS-BS channels
are first introduced. Then, we formulate the cascaded and UEs-IRS
channel estimation problems as two sparse signal recovery problems
in the frequency-domain.

\subsection{IRS Aided MISO Channel Model \label{subsec:Channel-Model}}

\emph{1. Channel model)} As mentioned above, the passive IRS aided
channel is split into two subchannels. Without loss of generality,
both $\boldsymbol{f}_{u}[k]$ and $\boldsymbol{G}[k]$ are presented
by the geometric channel model \cite{MA_IRS_THz}. Also, the channel
matrix $\boldsymbol{f}_{d,u}$ ($\boldsymbol{G}_{d}$) is assumed
to be frequency-selective, which is non-zero only for $d=1,\cdot\cdot\cdot,L_{\mathsf{c,f}}$
$\left(d=1,\cdot\cdot\cdot,L_{\mathsf{c,g}}\right)$ taps. We assume
that BS is equipped with Uniform Linear Arrays (ULAs) and the IRS
employs the $N_{\mathsf{i}}\left(N_{\mathsf{i}}=N_{\mathsf{i,x}}\times N_{\mathsf{i,y}}\right)$
uniform planar array (UPA). According to \cite{HAD_CE_TWC22}, the
discrete delay-domain channels of $d^{\mathsf{th}}$ delay tap are
given as
\begin{equation}
\boldsymbol{f}_{d,u}=\sqrt{\frac{N_{\mathsf{i}}}{L_{\mathsf{p,f}}}}\sum_{l=1}^{L_{\mathsf{p,f}}}\alpha_{l,u}p(dT_{\mathsf{s}}-\tau_{l,u})\boldsymbol{a}_{\mathsf{i}}(\theta_{\mathsf{i},l,u}^{\mathsf{A}},\varphi_{\mathsf{i},l,u}^{\mathsf{A}}),\label{eq:dd F}
\end{equation}
\begin{equation}
\boldsymbol{G}_{d}=\sqrt{\frac{N_{\mathsf{i}}N_{\mathsf{b}}}{L_{\mathsf{p,g}}}}\sum_{l=1}^{L_{\mathsf{p,g}}}\beta_{l}p(dT_{\mathsf{s}}-\tau_{l})\boldsymbol{a}_{\mathsf{b}}(\theta_{\mathsf{b},l}^{\mathsf{A}})\boldsymbol{a}_{\mathsf{i}}^{\mathsf{H}}(\theta_{\mathsf{i},l}^{\mathsf{D}},\varphi_{\mathsf{i},l}^{\mathsf{D}}),\label{eq:dd G}
\end{equation}
for the $u^{\mathsf{th}}$ user, $\alpha_{l,u}$ ($\beta_{l}$) and
$\tau_{l,u}$ denote propagation gain of the UEs-IRS (IRS-BS) channel
and delay corresponding to $l^{\mathsf{th}}$ path. $L_{\mathsf{p,f}}$
and $L_{\mathsf{p,g}}$ represent the number of $\boldsymbol{f}_{d,u}$
and $\boldsymbol{G}_{d}$ paths, respectively\textcolor{blue}{.} For
simplicity, we assume that the number of propagation paths for different
UEs remains constant; $p(\tau)$ is the pulse-shaping filter evaluated
at $\tau$ and $T_{s}$ denotes the sampling period; $\theta_{\mathsf{i},l,u}^{\mathsf{A}}$
($\varphi_{\mathsf{i},l,u}^{\mathsf{A}}$) and $\theta_{\mathsf{i},l}^{\mathsf{D}}$
($\varphi_{\mathsf{i},l}^{\mathsf{D}}$) represent azimuth (elevation)
angle of arrival (AoA) and angle of departure (AoD) of the $l^{\mathsf{th}}$
path at the IRS, respectively. $\theta_{\mathsf{b},l}^{\mathsf{A}}$
is the AoA of the $l^{\mathsf{th}}$ path at BS side. The array steering
vectors of the BS and the IRS are denoted by $\boldsymbol{a}_{\mathsf{b}}\in\mathbb{C}^{N_{\mathsf{b}}\times1}$
and $\boldsymbol{a}_{\mathsf{i}}\in\mathbb{C}^{N_{\mathsf{i}}\times1}$,
respectively. Thus the normalized array response vectors are expressed
as
\begin{equation}
\boldsymbol{a}_{\mathsf{b}}(\theta)=\frac{1}{\sqrt{N_{\mathsf{b}}}}\left[1\cdots e^{\frac{j2\pi d_{\mathsf{s}}}{\lambda_{\mathsf{c}}}(n_{\mathsf{b}}-1)\sin(\theta)}\cdots e^{\frac{j2\pi d_{\mathsf{s}}}{\lambda_{\mathsf{c}}}(N_{\mathsf{b}}-1)\sin(\theta)}\right]_{,}^{\mathsf{T}}
\end{equation}
\begin{equation}
\boldsymbol{a}_{\mathsf{i}}(\theta,\varphi)=\frac{1}{\sqrt{N_{\mathsf{i}}}}\left[1\cdots e^{\frac{j2\pi d_{\mathsf{s}}}{\lambda_{\mathsf{c}}}\left((n_{\mathsf{1}}-1)\sin(\theta)\cos(\varphi)+(n_{\mathsf{2}}-1)\sin(\varphi)\right)}\cdots e^{\frac{j2\pi d_{\mathsf{s}}}{\lambda_{\mathsf{c}}}\left((N_{\mathsf{i,x}}-1)\sin(\theta)\cos(\varphi)+(N_{\mathsf{i,y}}-1)\sin(\varphi)\right)}\right],
\end{equation}
where $d_{\mathsf{s}}$ and $\lambda_{\mathsf{c}}$ are the antenna
adjacent spacing and carrier wavelength, respectively, and we set
$d_{\mathsf{s}}/\lambda_{\mathsf{c}}=1/2$ without loss of generality.
During the estimating phase, both the BS and the user are considered
to know the array response vectors and IRS phase-shift matrices. Thus,
the channel can be reformulated more compactly as $\boldsymbol{f}_{d,u}=\boldsymbol{\boldsymbol{A}}_{\mathsf{R,i},u}\Delta_{\mathsf{f},d,u}\boldsymbol{a}_{\mathsf{T},u}^{\mathsf{H}}$,
where $\Delta_{\mathsf{f},d,u}\in\mathbb{C}^{L_{\mathsf{p,f}}\times L_{\mathsf{p,f}}}$
is diagonal with non-zero entries $\alpha_{l,u}p(dT_{\mathsf{s}}-\tau_{l,u})$,
which can be written as 
\begin{equation}
\Delta_{\mathsf{f},d,u}=\sqrt{\frac{N_{\mathsf{i}}}{L_{\mathsf{p,f}}}}\mathcal{\mathsf{diag}}\left(\alpha_{1,u}p(dT_{\mathsf{s}}-\tau_{1,u}),\ldots,\cdots\alpha_{L_{\mathsf{p,f}},u}p(dT_{\mathsf{s}}-\tau_{L_{\mathsf{p,f}},u})\right),\label{eq:Fd}
\end{equation}
where $\boldsymbol{\boldsymbol{A}}_{\mathsf{R,i},u}=\left[\boldsymbol{a}_{\mathsf{i}}(\theta_{\mathsf{i},1,u}^{\mathsf{A}},\varphi_{\mathsf{i},1,u}^{\mathsf{A}}),\ldots\boldsymbol{a}_{\mathsf{i}}(\theta_{\mathsf{i},L_{\mathsf{p,f}},u}^{\mathsf{A}},\varphi_{\mathsf{i},L_{\mathsf{p,f}},u}^{\mathsf{A}}),\right]\in\mathbb{C}^{N_{\mathsf{i}}\times L_{\mathsf{p,f}}}$
and $\boldsymbol{a}_{\mathsf{T},u}=\left[\theta_{1,u}^{\mathsf{D}},\ldots,\theta_{L_{\mathsf{p,f}},u}^{\mathsf{D}}\right]\in\mathbb{C}^{1\times L_{\mathsf{p,f}}}$
denote  the array steering vectors of AoAs at IRS and AoD at the $u^{\mathsf{th}}$
user side, respectively. Using (\ref{eq:dd F}) and (\ref{eq:dd G}),
the frequency-domain complex channel at subcarrier $k$ can be denoted
as
\begin{equation}
\boldsymbol{f}_{u}\mathit{\left[k\right]}=\sum_{d=1}^{L_{\mathsf{c,f}}}\boldsymbol{F}_{d,u}e^{-j\frac{2\pi dk}{K}}=\boldsymbol{\boldsymbol{A}}_{\mathsf{R,i},u}\mathsf{\text{\ensuremath{\mathbf{\Delta}_{\mathsf{f},u}}}\left[k\right]}\boldsymbol{a}_{\mathsf{T},u}^{\mathsf{H}},\label{eq:fd F}
\end{equation}
\begin{equation}
\boldsymbol{G}\left[k\right]=\sum_{d=1}^{L_{\mathsf{c,g}}}\boldsymbol{G}_{d}e^{-j\frac{2\pi dk}{K}}=\boldsymbol{A}_{\mathsf{R,b}}\mathsf{\text{\text{\ensuremath{\mathbf{\Delta}_{\mathsf{g}}}}}\mathit{\left[k\right]}}\boldsymbol{A}_{\mathsf{T,i}}^{\mathsf{H}},\label{eq:fd G}
\end{equation}
where $\text{\ensuremath{\mathbf{\Delta}_{\mathsf{f},u}}}\mathit{\left[k\right]}\in\mathbb{C}^{L_{\mathsf{c,f}}\times L_{\mathsf{c,f}}}$
and $\text{\ensuremath{\mathbf{\Delta}_{\mathsf{g}}}}\mathit{\left[k\right]}\in\mathbb{C}^{L_{\mathsf{c,g}}\times L_{\mathsf{c,g}}}$
are diagonal with non-zero complex entries such that $\text{\ensuremath{\mathbf{\Delta}_{\mathsf{f},u}}}\mathit{\left[k\right]}=$$\Sigma_{d=1}^{L_{\mathsf{c,f}}}\Delta_{\mathsf{f},d,u}e^{-j\frac{2\pi dk}{K}}$
and $\text{\text{\ensuremath{\mathbf{\Delta}_{\mathsf{g}}}}}\mathit{\left[k\right]}=\Sigma_{d=1}^{L_{\mathsf{c,g}}}\Delta_{\mathsf{g},d}e^{-j\frac{2\pi dk}{K}}$,
$k=1,2,\cdots K$.

\emph{2. Virtual channel model extention)} The training overhead of
cascaded channel estimation is proportional to the number of BS, IRS
and user antennas. Fortunately, due to the angular sparsity of the
mmWave  channel, we can formulate the cascaded channel estimation
as a sparse recovery problem \cite{Lin_TWC18} and adopt CS methods
to reduce pilot overhead. To mitigate the power leakage effect in
the angle domain, we further approximate the $\boldsymbol{f}_{u}\mathit{\left[k\right]}$
by designing redundant dictionary matrices with higher angular resolution.
The extended $\boldsymbol{f}_{u}\mathit{\left[k\right]}$ and $\boldsymbol{G}\mathit{\left[k\right]}$
in (\ref{eq:fd F}) and (\ref{eq:fd G}) are rewritten as
\begin{equation}
\boldsymbol{f}_{u}\mathit{\left[k\right]}=\widetilde{\boldsymbol{A}}_{\mathsf{R,i},u}\mathsf{\text{\ensuremath{\mathbf{\widetilde{\Delta}}_{\mathsf{f},u}}}\left[k\right]}\boldsymbol{\widetilde{a}}_{\mathsf{T},u}^{\mathsf{H}},\label{eq:ad F}
\end{equation}
\begin{equation}
\boldsymbol{G}\mathit{\left[k\right]}=\boldsymbol{\widetilde{A}}_{\mathsf{R,b}}\mathsf{\text{\ensuremath{\mathbf{\widetilde{\Delta}}_{\mathsf{g}}}}\mathit{\left[k\right]}}\boldsymbol{\widetilde{A}}_{\mathsf{T,i}}^{\mathsf{H}},\label{eq:ad G}
\end{equation}
where $\text{\ensuremath{\mathbf{\widetilde{\Delta}}_{\mathsf{f},u}}}\left[k\right]\in\mathbb{C}^{G_{\mathsf{i}}\times G_{\mathsf{u}}}$
and $\text{\ensuremath{\mathbf{\widetilde{\Delta}}_{\mathsf{g}}}}\mathit{\left[k\right]}\in\mathbb{C}^{G_{\mathsf{b}}\times G_{\mathsf{i}}}$
correspond to complex virtual angular-domain channels. The dictionary
matrices $\boldsymbol{\widetilde{a}}_{\mathsf{T},u}\in\mathbb{C}^{1\times G_{\mathsf{u}}}$
($\boldsymbol{\widetilde{A}}_{\mathsf{T,i}}\in\mathbb{C}^{N_{\mathsf{i}}\times G_{\mathsf{i}}}$)
and $\boldsymbol{\widetilde{A}}_{\mathsf{R,i},u}\in\mathbb{C}^{N_{\mathsf{i}}\times G_{\mathsf{i}}}$
($\boldsymbol{\widetilde{A}}_{\mathsf{R,b}}\in\mathbb{C}^{N_{\mathsf{b}}\times G_{\mathsf{b}}}$)
contain the UE (IRS) and IRS (BS) array response vectors evaluated
on a grid of size $G_{\mathsf{u}}$ ($G_{\mathsf{i}}$) and $G_{\mathsf{i}}$
($G_{\mathsf{b}}$) for the AoD and AoA, i.e., $\boldsymbol{\widetilde{a}}_{\mathsf{T},u}=\left[\theta_{u,1}^{\mathsf{D}},\ldots,\theta_{u,G_{\mathsf{u}}}^{\mathsf{D}}\right]$
and $\qquad\boldsymbol{\boldsymbol{\widetilde{A}}}_{\mathsf{R,i},u}=\left[\boldsymbol{a}_{\mathsf{i}}(\theta_{\mathsf{i},1,u}^{\mathsf{A}},\varphi_{\mathsf{i},1,u}^{\mathsf{A}}),\ldots\boldsymbol{a}_{\mathsf{i}}(\theta_{\mathsf{i},G_{\mathsf{i}},u}^{\mathsf{A}},\varphi_{\mathsf{i},G_{\mathsf{i}},u}^{\mathsf{A}})\right]$,
respectively. Due to the few number of scattering paths, we assume
that the grid of AoAs and AoDs of IRS have the same size $G_{\mathsf{i}}$. 

\subsection{Problem Formulation}

We assume that $U$ UEs use mutually orthogonal pilot signals \cite{tse2005comm},
the pilot signals of each user can be distinguished and processed
separately. Accordingly, by stacking $K$ UEs\textquoteright{} quantities,
the received signals of the $t^{\mathsf{th}}$ training symbol in
(\ref{eq:Y_U}) can be rewritten as
\begin{equation}
\boldsymbol{y}^{\left(t\right)}[k]=\boldsymbol{W}_{\mathsf{b}}^{\left(t\right)}[k]^{\mathsf{H}}\boldsymbol{H}^{(t)}[k]s^{\left(t\right)}[k]+\boldsymbol{n}_{\mathsf{c}}^{(t)}[k],\label{eq:Y_u}
\end{equation}
where $\boldsymbol{n}_{\mathsf{c}}^{(t)}[k]=\boldsymbol{W}_{\mathsf{b}}^{\left(t\right)}[k]^{\mathsf{H}}\boldsymbol{n}^{(t)}[k]$
is the combined noise vector received at the $k^{\mathsf{th}}$ subcarrier.

\emph{1) Cascaded channel with passive IRS:} In the $t^{\mathsf{th}}$
symbol training phase, the pilot symbol $s^{\left(t\right)}$ is known
at the receiver, and BS adopts the training combiner $\boldsymbol{W}_{\mathsf{b}}^{\left(t\right)}$.
Moreover, $s^{\left(t\right)}$, $\boldsymbol{W}_{\mathsf{b}}^{\left(t\right)}$
and $\boldsymbol{n}_{\mathsf{c}}^{(t)}$ are considered to be frequency-flat
to reduce the complexity of estimation \cite{SOMP18,HAD_CE_TWC22}.
Each entry in $\boldsymbol{W}_{\mathsf{b}}$ is normalized so that
their square modules are $\frac{1}{N_{\mathsf{b}}}$. We assume that
$\boldsymbol{f}\mathit{\left[k\right]}$ and $\boldsymbol{G}\left[k\right]$
are invariant for several consecutive symbols, and the frame duration
is shorter than the channel coherence time \cite{HAD_CE_TWC22}. Exploiting
the equality $\mathsf{vec}\left(\boldsymbol{AXB}\right)=\left(\boldsymbol{B}^{\mathsf{T}}\otimes\boldsymbol{A}\right)\mathsf{vec}\left(\boldsymbol{X}\right)$,
the received signals can be rewritten as
\begin{equation}
\boldsymbol{y}^{\left(t\right)}[k]=\left(s^{\left(t\right)\mathsf{T}}\otimes\boldsymbol{W}_{\mathsf{b}}^{(t)\mathsf{H}}\right)\mathsf{vec}\left(\boldsymbol{H}^{\left(t\right)}[k]\right)+\boldsymbol{n}_{\mathsf{c}}^{(t)}.\label{eq:vec_y}
\end{equation}
Combining (\ref{eq:ad F}) and (\ref{eq:ad G}) with (\ref{eq:cascaded channel}),
the vectorization of $\boldsymbol{H}^{\left(t\right)}[k]$ can be
expressed as
\begin{align}
 & \mathsf{vec}\left(\boldsymbol{H}^{\left(t\right)}[k]\right)\nonumber \\
 & =\mathsf{vec}\left(\boldsymbol{\widetilde{A}}_{\mathsf{R,b}}\widetilde{\boldsymbol{\Delta}}_{\mathsf{g}}\mathit{\left[k\right]}\boldsymbol{\widetilde{A}}_{\mathsf{T,i}}^{\mathsf{H}}\mathrm{diag}\left(\boldsymbol{r}^{\left(t\right)}\right)\boldsymbol{\widetilde{A}}_{\mathsf{R,i}}\mathsf{\mathsf{\text{\ensuremath{\mathbf{\widetilde{\Delta}}_{\mathsf{f}}}}\left[k\right]}}\boldsymbol{\widetilde{a}}_{\mathsf{T}}^{\mathsf{H}}\right)\nonumber \\
 & \overset{\left(a\right)}{=}\left(\boldsymbol{\widetilde{a}}_{\mathsf{T}}^{\mathsf{*}}\varotimes\boldsymbol{\widetilde{A}}_{\mathsf{R,b}}\right)\left(\boldsymbol{\widetilde{\Delta}}_{\mathsf{f}}^{\mathsf{T}}\mathit{\left[k\right]}\varotimes\widetilde{\boldsymbol{\Delta}}_{\mathsf{g}}\mathit{\left[k\right]}\right)\mathsf{vec}\left(\boldsymbol{\widetilde{A}}_{\mathsf{T,i}}^{\mathsf{H}}\mathrm{d}\left(\boldsymbol{r}^{\left(t\right)}\right)\boldsymbol{\widetilde{A}}_{\mathsf{R,i}}\right)\nonumber \\
 & \overset{\left(b\right)}{=}\left(\boldsymbol{\widetilde{a}}_{\mathsf{T}}^{\mathsf{*}}\varotimes\boldsymbol{\widetilde{A}}_{\mathsf{R,b}}\right)\left(\boldsymbol{\widetilde{\Delta}}_{\mathsf{f}}^{\mathsf{T}}\mathit{\left[k\right]}\varotimes\widetilde{\boldsymbol{\Delta}}_{\mathsf{g}}\mathit{\left[k\right]}\right)\boldsymbol{D}\boldsymbol{r}^{\left(t\right)}\nonumber \\
 & \overset{\left(c\right)}{=}\left(D_{\mathsf{g}}^{\mathsf{T}}\otimes\left(\boldsymbol{\widetilde{a}}_{\mathsf{T}}^{\mathsf{*}}\varotimes\boldsymbol{\widetilde{A}}_{\mathsf{R,b}}\right)\right)\widetilde{\boldsymbol{X}}\mathit{\left[k\right]}\boldsymbol{r}^{\left(t\right)},
\end{align}
where in $(a)$, $\left(\boldsymbol{\widetilde{\Delta}}_{\mathsf{f}}^{\mathsf{T}}\mathit{\left[k\right]}\varotimes\widetilde{\boldsymbol{\Delta}}_{\mathsf{g}}\mathit{\left[k\right]}\right)\in\mathbb{C}^{G_{\mathsf{u}}G_{\mathsf{b}}\times G_{\mathsf{i}}^{\mathsf{2}}}$
is a sparse vector of the virtual angular-domain channels with $L_{\mathsf{p,f}}L_{\mathsf{p,g}}$
non-zero elements, $(b)$ follows from the property $\mathsf{vec}(ABC)=(C^{\mathsf{T}}\odot A)b$,
where $b$ denotes the diagonal element vector of a diagonal matrix
and $\boldsymbol{D}\overset{\Delta}{=}\left(\boldsymbol{\widetilde{A}}_{\mathsf{R,i}}^{\mathsf{T}}\varodot\boldsymbol{\widetilde{A}}_{\mathsf{T,i}}^{\mathsf{H}}\right)\in\mathbb{C}^{G_{\mathsf{i}}^{\mathsf{2}}\times N_{\mathsf{i}}}$
is a sparse matrix. $(c)$ follows the guidance of \cite{wang2020compressed},
$\boldsymbol{D}$ can be further simplified. Specifically, the matrix
$\boldsymbol{D}$ only contains $G_{\mathsf{i}}$ distinct rows, which
are the first $G_{\mathsf{i}}$ rows of $\boldsymbol{D}$, i.e., $\boldsymbol{D}_{\mathsf{g}}\overset{\Delta}{=}\boldsymbol{D}(0:G_{\mathsf{i}}-1,:)$.
In addition, we define $\widetilde{\boldsymbol{X}}\mathit{\left[k\right]\in\mathbb{C}^{G_{\mathsf{u}}G_{\mathsf{b}}\times G_{\mathsf{i}}}}$
is a merged version of $\boldsymbol{Q}\overset{\Delta}{=}\boldsymbol{\widetilde{\Delta}}_{\mathsf{f}}^{\mathsf{T}}\mathit{\left[k\right]}\varotimes\widetilde{\boldsymbol{\Delta}}_{\mathsf{g}}\mathit{\left[k\right]}$,
i.e., $\widetilde{\boldsymbol{X}}\mathit{\left[k\right]}(:,i)=\sum_{j\epsilon M_{i}}\boldsymbol{Q}(:,j)$,
where $M_{i}$ is the set of indices corresponding to these rows in
$\boldsymbol{D}$. Leveraging these properties, the  $\boldsymbol{y}^{\left(t\right)}[k]$
in (\ref{eq:vec_y}) can be further represented as
\begin{align}
\boldsymbol{y}^{\left(t\right)}[k] & =\left(\boldsymbol{r}^{\left(t\right)\mathsf{T}}\otimes\left(s^{\left(t\right)\mathsf{T}}\otimes\boldsymbol{W}_{\mathsf{b}}^{(t)\mathsf{H}}\right)\right)\boldsymbol{\Psi}\boldsymbol{x}\mathit{\left[k\right]}+\boldsymbol{n}_{\mathsf{c}}^{(t)}[k]\nonumber \\
 & =\boldsymbol{\Phi}^{(t)}\boldsymbol{\Psi}\boldsymbol{x}\mathit{\left[k\right]}+\boldsymbol{n}_{\mathsf{c}}^{(t)}[k],\label{eq:Vec_Y}
\end{align}
where $\boldsymbol{\Psi}\overset{\Delta}{=}\boldsymbol{D}_{\mathsf{g}}^{\mathsf{T}}\varotimes\left(\boldsymbol{\widetilde{a}}_{\mathsf{T}}^{\mathsf{*}}\varotimes\boldsymbol{\widetilde{A}}_{\mathsf{R,b}}\right)\in\mathbb{C}^{N_{\mathsf{i}}N_{\mathsf{b}}\times G_{\mathsf{i}}G_{\mathsf{b}}G_{\mathsf{u}}}$
is the redundant dictionary of the cascaded channel; $\boldsymbol{\Phi}^{(t)}\overset{\Delta}{=}\left(\boldsymbol{r}^{\left(t\right)T}\otimes\left(s^{\left(t\right)\mathsf{T}}\otimes\boldsymbol{W}_{\mathsf{b}}^{(t)\mathsf{H}}\right)\right)$
denotes the measurement matrix and $\boldsymbol{x}\mathit{\left[k\right]}=\mathsf{vec}\left(\widetilde{\boldsymbol{X}}\mathit{\left[k\right]}\right)\in\mathbb{C}^{G_{\mathsf{u}}G_{\mathsf{b}}G_{\mathsf{i}}}$
is the sparse vector of the cascaded channel. By stacking $T$ successive
received pilots, we obtain the system model in the virtual angular-domain
as
\begin{equation}
\boldsymbol{y}[k]=\boldsymbol{\Phi\Psi}\boldsymbol{x}\mathit{\left[k\right]}+\boldsymbol{n}_{\mathsf{c}}[k],\label{eq:Vec_Y_t}
\end{equation}
where $\boldsymbol{y}\left[k\right]=\left[\boldsymbol{y}^{\left(1\right)}[k],\cdots,\boldsymbol{y}^{\left(T\right)}[k]\right]^{\mathsf{T}}\in\mathbb{C}^{M\times1}$
(\textbf{$M=T\times N_{\mathsf{s}}$}), $\boldsymbol{\Phi}=\left[\boldsymbol{\Phi}^{\left(1\right)},\cdots,\boldsymbol{\Phi}^{\left(T\right)}\right]^{\mathsf{T}}\in\mathbb{C}^{M\times N_{\mathsf{i}}N_{\mathsf{b}}}$
and $\boldsymbol{n}_{\mathsf{c}}[k]=\left[\boldsymbol{n}_{\mathsf{c}}^{\left(1\right)}[k],\cdots,\boldsymbol{n}_{\mathsf{c}}^{\left(T\right)}[k]\right]^{\mathsf{T}}$
of size $M\times1$. By stacking $K$ subcarriers quantities from
$\boldsymbol{y}[k]$, the vector $\boldsymbol{x}[k]$ in (\ref{eq:Vec_Y_t})
can be considered as a typical multiple-measurement-vectors (MMV)
sparse reconstruction problem, i.e.,
\begin{equation}
\mathsf{min}\left(\sum_{k=1}^{K}\left\Vert \boldsymbol{x}[k]\right\Vert _{1}\right),\quad\mathsf{s.t.}\left\Vert \boldsymbol{Y}-\boldsymbol{\Phi\Psi}\left(\boldsymbol{X}\right)\right\Vert _{2}^{2}<\varepsilon,\label{eq:problem}
\end{equation}
where $\boldsymbol{Y}=\left[\boldsymbol{y}[1],\cdots,\boldsymbol{y}[K]\right]\in\mathbb{C}^{M\times K}$,
$\boldsymbol{X}\in\mathbb{C}^{G_{\mathsf{i}}G_{\mathsf{b}}G_{\mathsf{u}}\times K}$
and $\varepsilon$ is error tolerance. 

\emph{2) Channel with hybrid IRS}: From the discussion in Section
\ref{subsec:Hybrid IRS}, the active elements receive orthogonal pilots
and estimate the UEs-IRS channel vectors. By leveraging tools from
CS and extending to $T_{\mathsf{i}}$ received signals, the uplink
signals at the IRS side can be formulated as
\begin{equation}
\boldsymbol{y}_{\mathsf{i}}[k]=\boldsymbol{\Phi}_{\mathsf{i}}\mathsf{vec}\left(\boldsymbol{\widetilde{A}}_{\mathsf{R,i}}\mathsf{\mathsf{\text{\ensuremath{\mathbf{\widetilde{\Delta}}_{\mathsf{i,f}}}}}}\left[k\right]\boldsymbol{\widetilde{a}}_{\mathsf{T}}^{\mathsf{H}}\right)+\boldsymbol{n}_{\mathsf{i}}[k]=\boldsymbol{\Phi}_{\mathsf{i}}\boldsymbol{\Psi}_{\mathsf{i}}\left(\boldsymbol{x}_{\mathsf{i}}[k]\right)+\boldsymbol{n}_{\mathsf{i}}[k]\text{.}\label{eq:vec_i_y}
\end{equation}
According to (\ref{eq:Vec_Y}) and (\ref{eq:Vec_Y_t}), $\boldsymbol{\Phi}_{\mathsf{i}}=s^{\mathsf{T}}\otimes\boldsymbol{w}_{\mathsf{i}}^{\mathsf{H}}\in\mathbb{C}^{T_{\mathsf{i}}\times N_{\mathsf{i}}}$
and $\boldsymbol{\Psi}_{\mathsf{i}}=\mathsf{vec}\left(\boldsymbol{\widetilde{a}}_{\mathsf{T}}^{\mathsf{*}}\varotimes\boldsymbol{\widetilde{A}}_{\mathsf{R,i}}\right)\in\mathbb{C}^{N_{\mathsf{i}}\times G_{\mathsf{u}}G_{\mathsf{i}}}$
denote the IRS reflecting measurement matrix and the redundant dictionary
of the UE-IRS channel, respectively. $\boldsymbol{n}_{\mathsf{i}}^{(t)}[k]$
is stacked noise vector of $k^{\mathsf{th}}$ subcarrier at the IRS
side and $\boldsymbol{x}_{\mathsf{i}}[k]=\mathsf{vec}\left(\mathsf{\mathsf{\text{\ensuremath{\mathbf{\widetilde{\Delta}}_{\mathsf{i,f}}}}}}\left[k\right]\right)$
represents the vectorization operation on the sparse vector of the
UE-IRS channel. We assume that the active elements are randomly selected
from the IRS reflectors \cite{CVDNCNN}. For simplicity, we employ
$G_{\mathsf{u}}=1$ in this work. 

There are various algorithms to solve sparse reconstruction MMV problem.
For example, the simultaneous weighted orthogonal matching pursuit
(SWOMP) algorithm \cite{SOMP18}, vector AMP (VAMP) technique \cite{Wu_VAMP}
and LAMP algorithm \cite{LAMP,MMVLAMP} neglect the sparse structure
of the sparse channel matrix and directly vectorize $\boldsymbol{y}^{\left(t\right)}[k]$
and the sparse channel matrix into vectors, which lead these CS algorithms
failing to achieve satisfactory accuracy in the IRS channels \cite{irs_struct}.
Meanwhile, some single measurement vector (SMV) \cite{SOMP18} and
MMV \cite{MMV20} algorithms assume that the channel has the same
sparse structure for all subcarriers, which ignores the information
in the frequency-domain and degrades the recovery performance \cite{Tensor_IRS}.
Therefore, these approaches motivate us to develop a channel estimator
for the IRS aided system by leveraging both sparse structure and frequency
properties.

\section{Proposed Hybrid Driven Networks for Channel Estimation }

Inspired by data driven and model driven networks, this section proposes
an attention network aided RLAMP network to solve the sparse channel
estimation problem. The network adopts a denoising convolution neural
network (DnCNN) to remove signal noise from the received signals.
By utilizing the higher resolution dictionary matrix, the attention
network extracts spatial information in the sparse channel and frequency
characteristics between different subcarriers. Next, the network structures
and training details of will be elaborated sequentially.

\subsection{Model Driven: Overview of the RLAMP Based Estimation Approach \label{subsec:RLAMP}}

\begin{figure}
\vspace{-0.6cm}

\centering\includegraphics[height=2in]{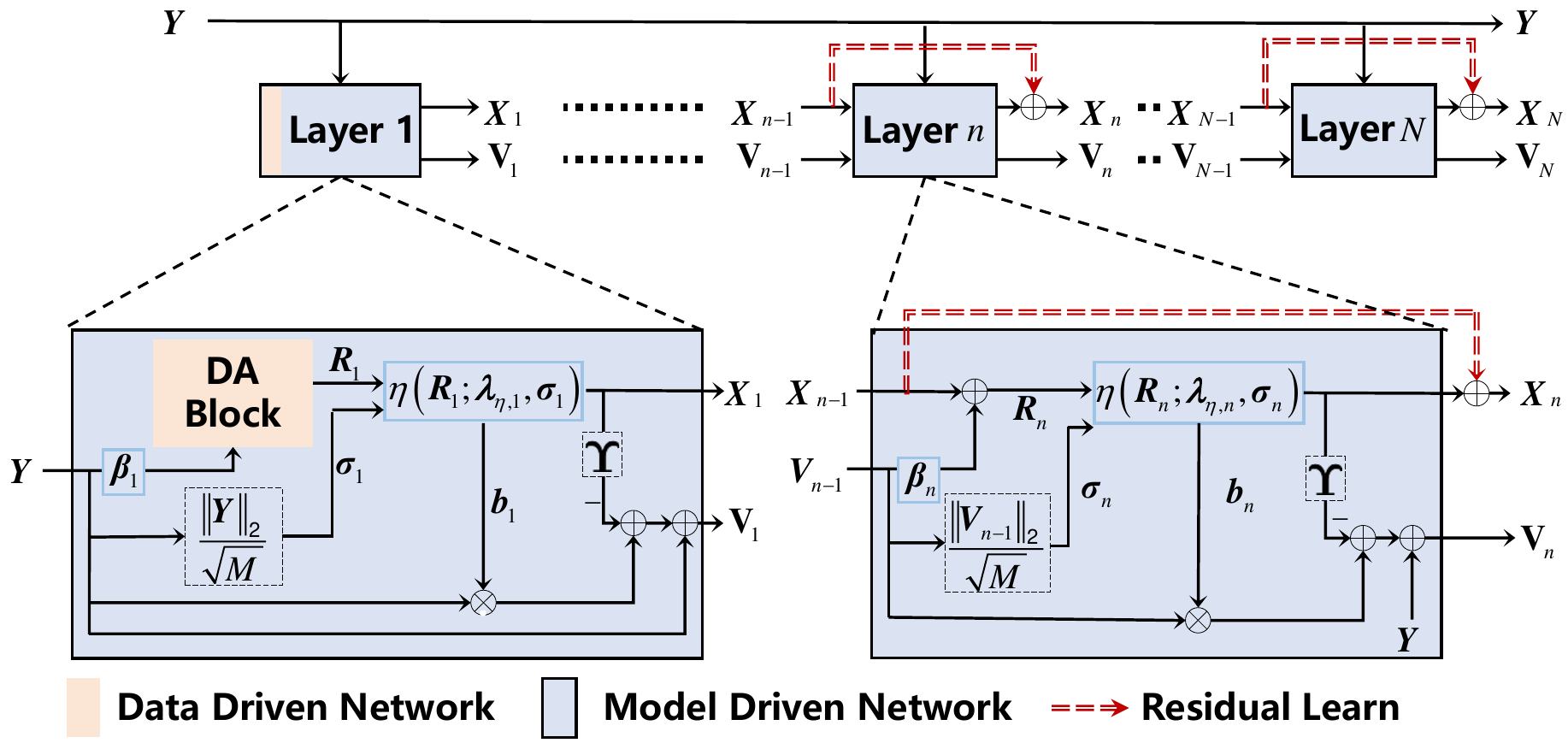}\caption{The $n^{\mathsf{th}}$ layer structure of the DA-RLAMP network.}

\vspace{-0.8cm}

\label{fig:LAMP structure }
\end{figure}

We integrate the LAMP network and the residual learning \cite{ResNet}
to the proposed networks. Before explicit estimation, the RLAMP network
requires to calculate the atom, which is defined as the vector in
the measurement matrix \cite{SOMP18}. The complex correlation vector
$\boldsymbol{C}\in\mathbb{C}^{G_{\mathsf{i}}G_{\mathsf{b}}\times K}$
is defined as
\begin{equation}
\left[\begin{array}{c}
\mathsf{Re}\left(\boldsymbol{\boldsymbol{C}}\right)\\
\mathsf{Im}\left(\boldsymbol{C}\right)
\end{array}\right]=\left[\begin{array}{c}
\mathsf{Re}\left(\boldsymbol{\beta}_{1}\right)\\
\mathsf{Im}\left(\boldsymbol{\beta}_{1}\right)
\end{array}\begin{array}{c}
-\mathsf{Im}\left(\boldsymbol{\beta}_{1}\right)\\
\mathsf{Re}\left(\boldsymbol{\beta}_{1}\right)
\end{array}\right]\left[\begin{array}{c}
\mathsf{Re}\left(\boldsymbol{Y}\right)\\
\mathsf{Im}\left(\boldsymbol{Y}\right)
\end{array}\right],\label{eq:C_matrix}
\end{equation}
where $\boldsymbol{\beta}_{1}$ is a trainable matrix with the same
size as matrix $\text{\ensuremath{\boldsymbol{\Upsilon}}}^{\mathsf{H}}$,
where $\boldsymbol{\boldsymbol{\Upsilon}}=\boldsymbol{\Phi\Psi}\in\mathbb{C}^{M\times G_{\mathsf{i}}G_{\mathsf{b}}}$
represents the equivalent measurement matrix. As Fig. \ref{fig:LAMP structure }
shows, the MMV-RLAMP extends the trainable parameters into the iteration
of the traditional AMP algorithms, which adaptively learn and optimize
the network from the channel samples. In LAMP \cite{modelLAMP}, the
input and output are typically limited to vectors, while the received
signals, measurement and estimated channel are matrices. Therefore,
the inputs and outputs at each iteration layer need to be vectorized.
Specifically, $\boldsymbol{X}_{n}\in\mathbb{C}^{G_{\mathsf{i}}G_{\mathsf{b}}\times K}$
and $\boldsymbol{V}_{n}\in\mathbb{C}^{M\times K}$ are the output
of the $n\mathsf{^{th}}$ layer ($n=1,\cdots N$). Following the guidance
of the AMP algorithm, the estimated matrix $\boldsymbol{\widehat{X}}_{n}$
and residual matrix $\boldsymbol{V}_{n}$ of each iteration in the
network are formulated as follows
\begin{equation}
\boldsymbol{\widehat{X}}_{n}=\eta\left(\boldsymbol{R}_{n};\boldsymbol{\lambda}_{\eta,n},\boldsymbol{\sigma}_{n}\right)+\boldsymbol{\widehat{X}}_{n-1},\label{eq:x_i}
\end{equation}
\begin{equation}
\boldsymbol{V}_{n}=\boldsymbol{Y}-\boldsymbol{\Upsilon}\boldsymbol{\widehat{X}}_{n}+\boldsymbol{b}_{n}\boldsymbol{V}_{n-1},\label{eq:V_i}
\end{equation}
where $\boldsymbol{R}_{1}=\boldsymbol{C}$, $\boldsymbol{V}_{0}=\boldsymbol{Y}$,
$\boldsymbol{\widehat{X}}_{0}=0$, and 
\begin{equation}
\boldsymbol{\sigma}_{n}=\frac{1}{\sqrt{M}}\left\Vert \boldsymbol{V}_{n-1}\right\Vert _{\mathsf{F}},\label{eq:sigma}
\end{equation}
\begin{equation}
\boldsymbol{b}_{n}=\frac{\lambda_{\mathsf{b},n}}{\sqrt{M}}\sum_{j=1}^{G}\frac{\partial\left[\eta\left(\boldsymbol{R}_{n};\boldsymbol{\lambda}_{\eta,n},\boldsymbol{\sigma}_{n}^{\mathsf{2}}\right)\right]_{j}}{\partial\left[\boldsymbol{R}_{n}\left(j,:\right)\right]},\label{eq:soft_th}
\end{equation}
\begin{equation}
\boldsymbol{R}_{n+1}=\boldsymbol{\widehat{X}}_{n}+\boldsymbol{\beta}_{n+1}\boldsymbol{V}_{n},\label{eq:R_i+1}
\end{equation}
with $G=G_{\mathsf{i}}G_{\mathsf{b}}$; the term $\boldsymbol{b}_{n}$
in (\ref{eq:V_i}) is the Onsager correction \cite{LAMP}, which is
derivative of $\eta(:,:)$ and accelerates the estimation convergence;
and $\boldsymbol{\beta}_{n}\in\mathbb{C}^{G_{\mathsf{i}}G_{\mathsf{b}}\times M}$
is a linear parameter of the $n\mathsf{^{th}}$ layer. Equation (\ref{eq:x_i})
indicates residual learning mechanism. Specifically, $\boldsymbol{\widehat{X}}_{n-1}$
is directly added to the outputs of the $n\mathsf{^{th}}$ layer.
This identity shortcut connection takes advantage of prior $n-1$
layers knowledge to further overcome the vanishing gradient problem
and accelerate training progress without introducing additional parameters
and computational complexity. The shrinkage function $\eta(:,:)$
in (\ref{eq:soft_th}) replaces the nonlinear activation function
in iteration, it can be expressed as
\begin{equation}
\left[\eta\left(\boldsymbol{R}_{n};\boldsymbol{\lambda}_{\eta,n},\boldsymbol{\sigma}_{n}^{\mathsf{2}}\right)\right]_{j}\overset{\Delta}{=}\lambda_{1\text{,}n}\mathsf{max}\left(\left|\boldsymbol{r}_{n,j}\right|-\lambda_{2,n}\boldsymbol{\sigma}_{n}\text{,0}\right)\mathsf{sgn}\text{(\ensuremath{\boldsymbol{r}_{n,j}})},
\end{equation}
where $\boldsymbol{r}_{n,j}=\boldsymbol{R}_{n}(j,:)$ express $j^{\mathsf{th}}$
row of $\boldsymbol{R}_{n}$; and $\boldsymbol{\lambda}_{n}=\left\{ \lambda_{1\text{,}n},\lambda_{2\text{,}n},\lambda_{\mathsf{b},n}\right\} \in\mathbb{R}^{3}$
is the predefined and nonlinear shrinkage parameter of the $n\mathsf{^{th}}$
layer and $\boldsymbol{\lambda}_{\eta,n}=\left\{ \lambda_{1\text{,}n},\lambda_{2\text{,}n}\right\} \in\mathbb{R}^{2}$.
The linear parameter $\boldsymbol{\beta}_{n}$ and nonlinear shrinkage
parameter $\boldsymbol{\lambda}_{n}$ are regarded as trainable parameters,
which are learned in each iteration from training data. Roughly speaking,
our network adopts a soft-Threshold function rather than other denoiser
such as denoising-based LAMP (LDAMP) \cite{LDAMP} and Gaussian mixture
LAMP (GM-LAMP) \cite{GMAMP}. We propose a denoising network before
the RLAMP, which learns and mitigates the noise component in the $\boldsymbol{x}[k]$,
the details described in subsection \ref{subsec:denoiser and attention }.

However, the MMV-LAMP networks for channel estimation suffer from
the following problems: (i) The traditional LAMP network and its variants
are applicable for any signal without considering the channel sparse
structure; (ii) Some developed architectures focus on the design of
denoiser (shrinkage function) \cite{GMAMP,HN-LAMP} to obtain residual
noise vector in different iterations, which neglect the channel properties.
To improve the estimation performance, we will propose an MMV-DA-RLAMP
network based on a denoising network and the attention mechanism in
the next subsection. The proposed network implicitly learns the sparse
characteristics of signals without any channel prior information.

\subsection{Data Driven: Channel Feature Filtering and Enhancement Via Denoising
Network and Attention Mechanism \label{subsec:denoiser and attention }}

\begin{figure}
\vspace{-0.6cm}

\centering\includegraphics[height=2in]{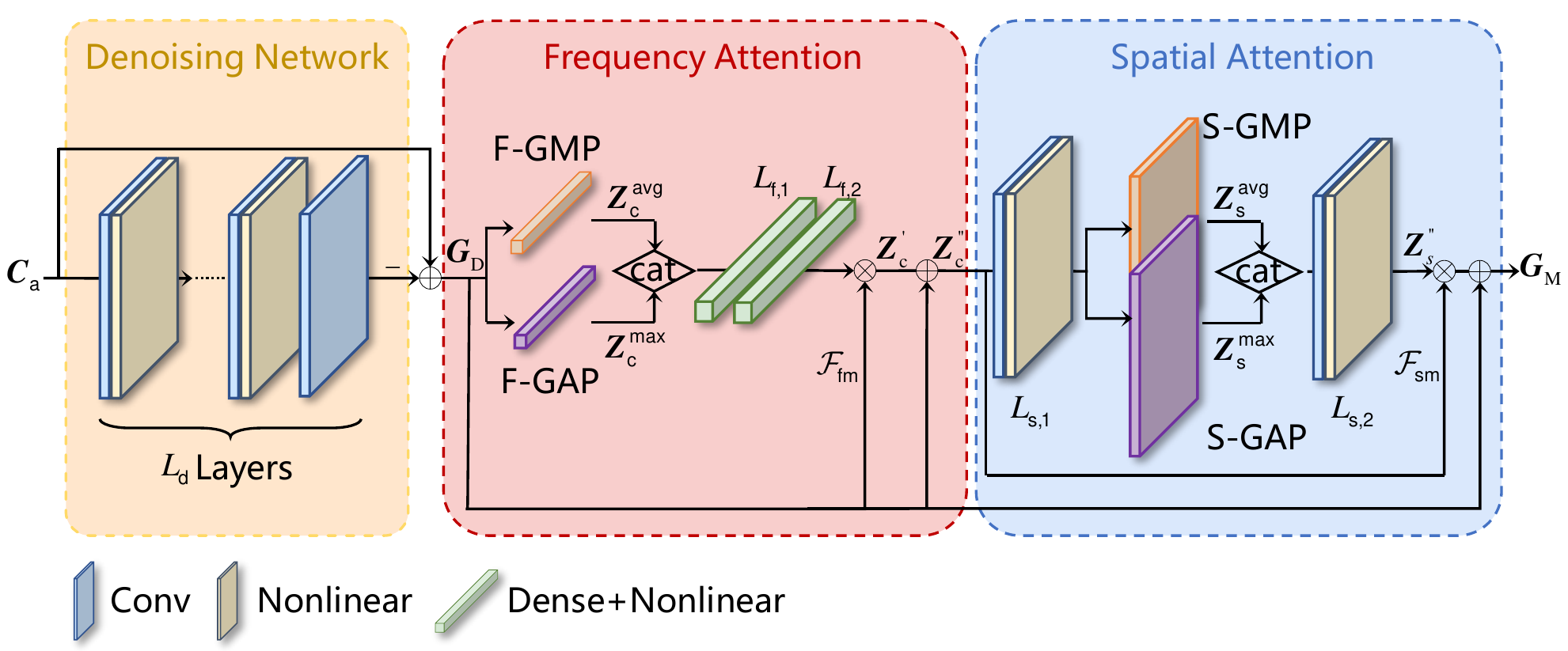}
\caption{The proposed denoising network and attention mechanism (DA block)
for channel amplitude estimation and feature enhancement.}

\vspace{-0.8cm}

\label{fig:DnCNN=000026Att}
\end{figure}

In Fig. \ref{fig:DnCNN=000026Att}, the denoising network consists
of $L_{\mathsf{d}}$ denoising layers to learn the residual noise
from the noisy correlation matrix $\boldsymbol{C}$. Since $\boldsymbol{\boldsymbol{\Upsilon}}$
is a complex-valued matrix, we divide the real part and imaginary
part to facilitate feature extraction. Table \ref{table:Hyperparameters}
summarizes the hyperparameters of DnCNN and the attention network.

\begin{table}[h]
\caption{Hyperparameters of the DnCNN and Attention Network.}

\renewcommand\arraystretch{0.6}

\centering%
\begin{tabular}{ccc}
\hline 
\multicolumn{3}{l}{\textbf{Input}: Noisy correlation vector $\boldsymbol{C}_{\mathsf{a}}\in\mathbb{R}^{2\times G_{\mathsf{i}}\times G_{\mathsf{b}}\times K}$}\tabularnewline
\hline 
\multicolumn{3}{c}{\textbf{Denoising Network}}\tabularnewline
\hline 
Layers & Operations & Filter Size\tabularnewline
1\textasciitilde$L_{\mathsf{d}}-$1 & Conv+Nonlinear & $K\times\left(3\times3\times K\right)$\tabularnewline
$L_{\mathsf{d}}$ & Conv & $K\times\left(3\times3\times K\right)$\tabularnewline
\hline 
\multicolumn{3}{c}{\textbf{Attention network}}\tabularnewline
\hline 
Layers & Operations & Size\tabularnewline
$L_{\mathsf{f,1}}$ & Dense+Nonlinear & $2K$\tabularnewline
$L_{\mathsf{f,2}}$ & Dense+Nonlinear & $K$\tabularnewline
$L_{\mathsf{s,1}}$ & Conv+Nonlinear & $K\times\left(3\times3\times K\right)$\tabularnewline
$L_{\mathsf{s,2}}$ & Conv+Nonlinear & $1\times\left(3\times3\times2\right)$\tabularnewline
\hline 
\multicolumn{3}{l}{\textbf{Output}: Channel frequency--spatial feature $\boldsymbol{G}_{\mathsf{M}}\in\mathbb{C}^{G_{\mathsf{i}}\times G_{\mathsf{b}}\times K}$}\tabularnewline
\hline 
\end{tabular}

\vspace{-0.8cm}

\label{table:Hyperparameters}
\end{table}
 Since the sparsity of the real part and imaginary part of the sparse
matrix is the same, and they are generally orthogonal, we present
two parallel images for the real and imaginary part of the noisy input,
i.e., $\boldsymbol{C}_{\mathsf{a}}=\mathcal{F}_{\mathsf{RV}}\left(\boldsymbol{C}^{'}\right)$,
where $\boldsymbol{C}_{\mathsf{a}}\in\mathbb{R}^{2\times G_{\mathsf{i}}\times G_{\mathsf{b}}\times K}$
is the input of the DnCNN, $\mathcal{F}_{\mathsf{RV}}(\cdot):\mathbb{C}^{G_{\mathsf{i}}\times G_{\mathsf{b}}\times K}\mapsto\mathbb{R}^{2\times G_{\mathsf{i}}\times G_{\mathsf{b}}\times K}$
denotes a mapping function which builds a real-valued matrix based
on a complex-valued matrix, and $\boldsymbol{C}^{'}[k]\in\mathbb{C}^{G_{\mathsf{i}}\times G_{\mathsf{b}}}$
denotes the $k^{\mathsf{th}}$ column of $\boldsymbol{C}^{'}$
\begin{equation}
\boldsymbol{C}^{'}[k]=\mathsf{vec2mat}\left(\boldsymbol{c}[k],[G_{\mathsf{i}}\times G_{\mathsf{b}}]\right),\forall k.\label{eq:vec_c}
\end{equation}
As shown in Fig. \ref{fig:DnCNN=000026Att}, $L_{\mathsf{d}}$ convolution
(Conv) layers to gradually extract the noise characteristics. Specifically,
the denoiser is composed of a residual block \cite{ResNet} and a
subtract operator. The Conv operation and nonlinear activation function
(Nonlinear) are utilized to obtain the noise characteristics of the
correlation matrix at the first $L_{\mathsf{d}}-1$ layer, where the
nonlinear activation function leverages Leaky correction linear unit
(LeakyReLU) to solve the dying ReLU problem of the negative part of
the input \cite{Leakyrelu}. The final layer applies a convolution
layer to reconstruct the residual noise map without activation function.
The residual noise in $\boldsymbol{C}_{\mathsf{a}}$ is additive,
and the DnCNN \cite{DnCNN} aims to learn the residual map $\mathcal{F}_{\mathsf{D}}(\boldsymbol{C}_{\mathsf{a}};\boldsymbol{\Theta}_{\mathsf{d}})$
rather than denoise directly, where $\mathcal{F}_{\mathsf{D}}(\cdot;\boldsymbol{\Theta}_{\mathsf{d}})$
denotes the function expression of DnCNN with the network parameters
$\boldsymbol{\Theta}_{\mathsf{d}}=\left[\boldsymbol{\Theta}_{\mathsf{d,0}},\cdots\boldsymbol{\Theta}_{\mathsf{d,L_{\mathsf{d}}-1}}\right]$
\footnote{DnCNN in \cite{DnCNN} is not our innovation, and thereby, detailed
explicit expression is omitted here due to space limitation.}. Finally, the element-wise subtraction operation is adopted between
the input matrix and output of the DnCNN networks to derive a noisy-clean
matrix $\boldsymbol{G}_{\mathsf{D}}$, which can be expressed as
\begin{equation}
\boldsymbol{G}_{\mathsf{D}}=\mathcal{F}_{\mathsf{CV}}\left(\boldsymbol{C}_{\mathsf{a}}-\mathcal{F}_{\mathsf{D}}(\boldsymbol{C}_{\mathsf{a}};\boldsymbol{\Theta})\right)\in\mathbb{C}^{G_{\mathsf{i}}\times G_{\mathsf{b}}\times K},\label{eq:Dncnn_map}
\end{equation}
where $\mathcal{F}_{\mathsf{CV}}(\cdot):\mathbb{R}^{2\times G_{\mathsf{i}}\times G_{\mathsf{b}}\times K}\mapsto\mathbb{C}^{G_{\mathsf{i}}\times G_{\mathsf{b}}\times K}$
denotes the complex-valued mapping function. 

Similar to the attention mechanism and structure in \cite{2018cbam},
we propose a frequency and spatial attention network (FSAN) to enhance
both spatial sparse structure and frequency features simultaneously.
Fig.  \ref{fig:DnCNN=000026Att} shows the overall structure of the
FSAN, which is composed of a frequency attention network (FAN) and
a spatial attention network (SAN). The FAN takes the matrix $\boldsymbol{G}_{\mathsf{D}}$
as the input, and the input of the SAN requires the output of FAN
and $\boldsymbol{G}_{\mathsf{D}}$. The above-mentioned two attention
networks correspond to the subcarrier feature selection and the spatial
feature selection in estimation processing, respectively.

1.\emph{ FAN)} The structural information of the received signals
in the frequency-domain has been considered in wideband massive MIMO-OFDM
systems \cite{Tensor_IRS}. Furthermore, the frequency characteristics
are difficult to be characterized by conventional approaches. Conversely,
attention mechanisms have the powerful capability to focus on relevant
information from the data. As Fig. \ref{fig:DnCNN=000026Att} shows,
the frequency attention network maps the $\boldsymbol{G}_{\mathsf{D}}$
to a reweighted frequency feature vector, which implies the information
features in frequency-domain and generates a weighting factor for
each subcarrier. Each subcarrier of $\boldsymbol{G}_{\mathsf{D}}$
is squeezed into a single numeric value using frequency-wise global
average pooling (F-GAP) and frequency-wise global max pooling (F-GMP),
which are calculated by
\[
\boldsymbol{z}_{\mathsf{c}}^{\mathsf{avg}}[k]=\frac{1}{G_{\mathsf{i}}\times G_{\mathsf{b}}}\sum_{i=1}^{G_{\mathsf{i}}}\sum_{j=1}^{G_{\mathsf{b}}}\mathcal{F}_{\mathsf{RV}}\left(g_{\mathsf{D}}^{i,j}[k]\right),\forall k,
\]
\begin{equation}
\boldsymbol{z}_{\mathsf{c}}^{\mathsf{max}}[k]=\mathsf{max}\left(\mathcal{F}_{\mathsf{RV}}\left(\boldsymbol{G}{}_{\mathsf{D}}[k]\right)\right),\forall k,\label{eq:CA_op}
\end{equation}
where $\boldsymbol{z}_{\mathsf{c}}^{\mathsf{avg}}[k]\in\mathbb{R}^{2}$
and $\boldsymbol{z}_{\mathsf{c}}^{\mathsf{max}}[k]\in\mathbb{R}^{2}$
denote the average and maximum values of the real and imaginary parts
of $\boldsymbol{G}_{\mathsf{D}}[k]$, respectively; and  $g_{\mathsf{D}}^{i,j}[k]$
is the value at position $(i,j)$ of $\boldsymbol{G}_{\mathsf{D}}[k]$
\footnote{Similar to the above DnCNN, two attention networks divide the real
and imaginary part of the input and enhance the features of the two
parallel matrices, respectively.}. According to Table \ref{table:Hyperparameters}, two fully connected
(FC) layers with LeakyReLU activation function is operated as $\boldsymbol{z}_{\mathsf{c}}[k]=\mathcal{F}_{\mathsf{fc}}(\boldsymbol{z}_{\mathsf{c}}^{\mathsf{avg}}[k],\boldsymbol{z}_{\mathsf{c}}^{\mathsf{max}}[k])$,
where $\mathcal{F}_{\mathsf{fc}}$ is the frequency-wise concatenation
operation. In short, the frequency attention map of each subcarrier
is computed as
\begin{equation}
\boldsymbol{z}_{\mathsf{c}}^{'}[k]=\mathcal{F}_{\mathsf{Fc}}\left(\mathcal{F}_{\mathsf{Fc}}\left(\boldsymbol{z}_{\mathsf{c}}[k];\boldsymbol{\Theta}_{\mathsf{f,1}}\right);\boldsymbol{\Theta}_{\mathsf{f\text{,2}}}\right)\in\mathbb{R}^{2},\label{eq:CA_FC}
\end{equation}
where $\mathcal{F}_{\mathsf{Fc}}$ and $\boldsymbol{\Theta}_{\mathsf{f}}=\left[\boldsymbol{\Theta}_{\mathsf{f,1}},\boldsymbol{\Theta}_{\mathsf{f\text{,2}}}\right]$
refer to the FC layers and corresponding parameters, respectively.
A skip connection is design from the input to the output directly
to learn the residual and fast convergence. Thus the final output
of the frequency attention network $\boldsymbol{Z}_{\mathsf{c}}^{''}[k]\in\mathbb{R}^{2\times G_{\mathsf{i}}\times G_{\mathsf{b}}}$
is obtained by 
\begin{equation}
\boldsymbol{Z}_{\mathsf{c}}^{''}[k]=\mathcal{F}_{\mathsf{RV}}\left(\boldsymbol{G}_{\mathsf{D}}\right)+\mathcal{F}_{\mathsf{fm}}\left(\mathcal{F}_{\mathsf{RV}}\left(\boldsymbol{G}_{\mathsf{D}}\right),\boldsymbol{z}_{\mathsf{c}}^{'}[k]\right),\forall k,\label{eq:CA_out}
\end{equation}
and $\mathcal{F}_{\mathsf{fm}}(\cdot,\cdot)$ refers to frequency-wise
multiplication between the input and frequency feature map.

2.\emph{ SAN)} SAN adaptively emphasizes the informative features
in the virtual angular-domain, e.g., it enhances the non-zero elements
and their sparse structure, and weakens less informative spatial information
(zero elements). Fig. \ref{fig:DnCNN=000026Att} shows the spatial-wise
GMP and GAP, i.e., S-GMP and S-GAP operate on the input matrix $\boldsymbol{Z}_{\mathsf{s}}=\varphi\left(\mathcal{F}_{\mathsf{conv}}\left(\boldsymbol{Z}_{\mathsf{c}}^{''};\boldsymbol{\Theta}_{\mathsf{s,1}}\right)\right)$,
which can be expressed as 
\begin{align}
\boldsymbol{Z}_{\mathsf{s}}^{\mathsf{avg}} & =\frac{1}{K}\sum_{k=1}^{K}\boldsymbol{Z}_{\mathsf{s}}[k]\in\mathbb{R}^{2\times G_{\mathsf{i}}\times G_{\mathsf{b}}},\\
\boldsymbol{Z}_{\mathsf{s}}^{\mathsf{max}} & =\mathsf{max}\left(\boldsymbol{Z}_{\mathsf{s}}[k]\right)\in\mathbb{R}^{2\times G_{\mathsf{i}}\times G_{\mathsf{b}}},\label{eq:SA_op}
\end{align}
where $\varphi$ and $\mathcal{F}_{\mathsf{conv}}$ denote activation
function and convolutional layer, respectively; and $\boldsymbol{\Theta}_{\mathsf{s}}=[\boldsymbol{\Theta}_{\mathsf{s},1},\boldsymbol{\Theta}_{\mathsf{s,2}}]$
is the network parameters of the spatial attention network. Then two
outputs are concatenated as the input of a new convolutional layer,
which is calculated by
\begin{equation}
\boldsymbol{Z}_{\mathsf{s}}^{''}=\varphi\left(\mathcal{F}_{\mathsf{conv}}\left(\boldsymbol{Z}_{\mathsf{s}}^{'};\boldsymbol{\Theta}_{\mathsf{s,1}}\right)\right)\in\mathbb{R}^{2\times G_{\mathsf{i}}\times G_{\mathsf{b}}},\label{eq:SA_CNN}
\end{equation}
where $\boldsymbol{Z}_{\mathsf{s}}^{'}=\mathcal{F}_{\mathsf{sc}}(\boldsymbol{Z}_{\mathsf{s}}^{\mathsf{avg}},\boldsymbol{Z}_{\mathsf{s}}^{\mathsf{max}})\in\mathbb{R}^{2\times G_{\mathsf{i}}\times G_{\mathsf{b}}\times2}$
denotes the output of spatial-wise concatenation operation $\mathcal{F}_{\mathsf{sc}}$.
The frequency attention map $\boldsymbol{Z}_{\mathsf{c}}^{''}$ and
sparse attention map $\boldsymbol{Z}_{\mathsf{s}}^{''}$ are spatial-wise
multiplied to scale the feature maps adaptively. Note that a global
skip connection is used to learn the residual between the $\boldsymbol{G}_{\mathsf{D}}$
and attention network. Hence the frequency--spatial feature of the
$k^{\mathsf{th}}$ subcarrier are obtained according to
\begin{equation}
\boldsymbol{G}_{\mathsf{M}}[k]=\boldsymbol{G}_{\mathsf{D}}[k]+\mathcal{F}_{\mathsf{CV}}\left(\mathcal{F}_{\mathsf{sm}}\left(\boldsymbol{Z}_{\mathsf{s}}^{''},\boldsymbol{Z}_{\mathsf{c}}^{''}[k]\right)\right),\forall k,\label{eq:CA_SA}
\end{equation}
where $\mathcal{F}_{\mathsf{sm}}(\cdot,\cdot)$ denotes the spatial-wise
multiplication between the input and the spatial feature map.
\begin{rem}
Note that the denoising network and attention network have strong
scalability \cite{DLbook}. Although the size of $\boldsymbol{\boldsymbol{\Upsilon}}$
changes with $\boldsymbol{\Phi}$ and $\boldsymbol{\Psi}$, the input
and output of the network are $G_{\mathsf{i}}\times G_{\mathsf{b}}\times K$
and $G_{\mathsf{i}}G_{\mathsf{b}}\times K$ matrices, which can easily
change with the size of the measurements. Also, the number of filters
depends on $K$.
\end{rem}

\subsection{Proposed Hybrid Driven Network: DA-RLAMP Network for Passive IRS
Aided Channel Estimation}

Based on the above analysis, we summarize the hybrid driven network
based channel estimation scheme as Algorithm \ref{alg:DA-RLAMP} and
Fig. \ref{fig:LAMP structure }. It is clear that DA-RLAMP is structured
based on three main procedures after the initialization steps in line
1.
\begin{algorithm}
\caption{DA-RLAMP Algorithm}

\begin{algorithmic}[1]

\Require\textbf{ }The received pilots $\boldsymbol{Y}$, the measurement
matrix $\boldsymbol{\Phi}$, the redundant dictionary $\boldsymbol{\Psi}$,
the number of iterations $N$, trainable parameters $\boldsymbol{\Theta}$,
$\boldsymbol{\beta}$ and $\boldsymbol{\lambda}$.

\State $\boldsymbol{C}[k]\shortleftarrow\boldsymbol{\beta}_{0}\boldsymbol{Y}$,
$\forall k$

\State $\boldsymbol{G}_{\mathsf{M}}[k]$ $\shortleftarrow$ Feature
Enhancement ($\boldsymbol{\Theta}$, $\boldsymbol{C}[k]$, $\forall k$)

\State $\widehat{\boldsymbol{X}}$ $\shortleftarrow$ Channel Estimation
($\boldsymbol{\boldsymbol{\Upsilon}}$, $\boldsymbol{Y}$, $N$, $\boldsymbol{G}_{\mathsf{M}}$,
$\boldsymbol{\beta}$, $\boldsymbol{\lambda}$)

\State $\boldsymbol{\widehat{H}}_{\mathsf{c}}$ $\shortleftarrow$
Channel Reconstruction ($\widehat{\boldsymbol{X}}$, $\boldsymbol{\Psi}$)

\Ensure $\boldsymbol{\widehat{H}}_{\mathsf{c}}$

\Procedure{DnCNN and Attention Aided Sparse Enhancement Feature Enhancement}{$\boldsymbol{\Theta}, \boldsymbol{C}[k], \forall k$}

\State$\boldsymbol{C}_{\mathsf{m}}[k]$=$\mathsf{vec2mat}\left(\boldsymbol{C}[k],[G_{\mathsf{i}}\times G_{\mathsf{b}}]\right)$
// per (\ref{eq:vec_c})

\State$\boldsymbol{G}_{\mathsf{D}}$$\overset{\mathsf{DnCNN}}{\longleftarrow}\{\boldsymbol{C}^{'};\boldsymbol{\Theta}_{\mathsf{d}}\}$//
cf., Fig. \ref{fig:DnCNN=000026Att}

\State$\boldsymbol{Z}_{\mathsf{c}}^{''}[k]$$\overset{\mathsf{FAN}}{\longleftarrow}$$\{\boldsymbol{G}_{\mathsf{D}};\boldsymbol{\Theta}_{\mathsf{f}}\}$

\State$\boldsymbol{Z}_{\mathsf{s}}^{'}\overset{\mathsf{SAN}}{\longleftarrow}\{\boldsymbol{Z}_{\mathsf{c}}^{''}[k];\boldsymbol{\Theta}_{\mathsf{s}}\}$,
$\forall k$

\State$\boldsymbol{G}_{\mathsf{M}}[k]=\boldsymbol{G}_{\mathsf{D}}+\mathcal{F}_{\mathsf{CV}}\left(\mathcal{F}_{\mathsf{sw}}\left(\boldsymbol{Z}_{\mathsf{s}}^{'},\boldsymbol{Z}_{\mathsf{c}}^{''}[k]\right)\right)$,
$\forall k$ //per (\ref{eq:CA_SA})

\Statex\textbf{return} $\boldsymbol{G}_{\mathsf{M}}$

\EndProcedure

\Procedure{Channel Estimation Based on RLAMP}{$\boldsymbol{\boldsymbol{\Upsilon}}, \boldsymbol{Y}, N, \boldsymbol{G}_{\mathsf{M}}, \boldsymbol{\beta}, \boldsymbol{\lambda}$}

\State Initialization $\boldsymbol{V}_{0}=\boldsymbol{Y}$, $\widehat{\boldsymbol{\boldsymbol{X}}}_{0}=0$,
$\boldsymbol{R}_{1}=\boldsymbol{G}_{\mathsf{M}}$

\For{$n=1,\cdots,N$}

\State$\boldsymbol{\sigma}_{n}=\frac{1}{\sqrt{M}}\left\Vert \boldsymbol{V}_{n-1}\right\Vert _{\mathsf{F}}$
// per (\ref{eq:sigma})

\State$\boldsymbol{\widehat{X}}_{n}=\eta\left(\boldsymbol{R}_{n};\boldsymbol{\lambda}_{\eta,n},\boldsymbol{\sigma}_{n}\right)+\boldsymbol{\widehat{X}}_{n-1}$
// per (\ref{eq:x_i})

\State$\boldsymbol{b}_{n}=\frac{\lambda_{\mathsf{b},n}}{\sqrt{M}}\sum_{j=1}^{G}\frac{\partial\left[\eta\left(\boldsymbol{R}_{n};\boldsymbol{\lambda}_{\eta,n},\boldsymbol{\sigma}_{n}^{\mathsf{2}}\right)\right]_{j}}{\partial\left[\boldsymbol{R}_{n}\left(j,:\right)\right]},$
// per (\ref{eq:soft_th})

\State$\boldsymbol{V}_{n}=\boldsymbol{Y}-\boldsymbol{\Upsilon}\boldsymbol{\widehat{X}}_{n}+\boldsymbol{b}_{n}\boldsymbol{V}_{n-1}$
// per (\ref{eq:V_i})

\State$\boldsymbol{R}_{n+1}=\boldsymbol{\widehat{X}}_{n}+\boldsymbol{\beta}_{n+1}\boldsymbol{V}_{n}$
// per (\ref{eq:R_i+1})

\EndFor

\Statex\textbf{return} $\boldsymbol{\widehat{X}}_{N}$

\EndProcedure

\Procedure{Frequency-domain Channel Reconstruction}{$\boldsymbol{\widehat{X}}_{N}, \boldsymbol{\Psi}$}

\For{$k=1,\cdots,K$}

\State$\mathsf{vec}\left(\boldsymbol{\widehat{H}}_{\mathsf{c}}\mathit{\left[k\right]}\right)=\boldsymbol{\Psi}\mathsf{vec}\left(\boldsymbol{\widehat{X}}_{N}[k]\right)$

\EndFor

\Statex \textbf{return} $\boldsymbol{\widehat{H}}_{\mathsf{c}}$

\EndProcedure

\end{algorithmic}

\label{alg:DA-RLAMP}
\end{algorithm}

\emph{1) DnCNN and Attention Aided Feature Enhancement (lines 5-11):}
Given the trained parameters denoted as $\boldsymbol{\Theta}=\left\{ \boldsymbol{\Theta}_{\mathsf{d}},\boldsymbol{\Theta}_{\mathsf{f}},\boldsymbol{\Theta}_{\mathsf{s}}\right\} $,
lines 6 and 7 of the proposed network first compute DnCNN input $\boldsymbol{c}_{\mathsf{m}}[k]$
by mapping $K$ vectors into a correlation matrix form as per (\ref{eq:vec_c})
and then obtain the DnCNN output $\boldsymbol{G}_{\mathsf{D}}$. The
attention procedure consists of frequency and spatial parts as shown
in lines 8 and 9 of Algorithm \ref{alg:DA-RLAMP}, respectively. These
steps are explained in detail in Section \ref{subsec:denoiser and attention }.

\emph{2) Channel Estimation Based on RLAMP (lines 12-21):} After the
initialization step in line 13, the proposed RLAMP architecture estimates
$X$ and the residual by iteratively minimizing the estimation error.
Meanwhile, the trainable linear transform measurement matrices $\boldsymbol{\beta}$
and nonlinear shrinkage coefficients $\boldsymbol{\lambda}$ are optimized
by backpropagation in the training phase. Inspired by deep complex
networks \cite{cv_network}, the complex-valued $\boldsymbol{\beta}_{n}$
of each iteration consists of two real-valued matrices corresponding
to $\mathsf{Re}\left\{ \boldsymbol{\beta}_{n}\right\} $ and $\mathsf{Im}\left\{ \boldsymbol{\beta}_{n}\right\} $.
More details are depicted in Section \ref{subsec:RLAMP}.

\emph{3) Wideband Frequency-domain Channel Reconstruction (lines 22-26):}
Once all the supports and channel gains of the virtual angular-domain
channel are estimated, the final estimated frequency-domain channel
matrix of the $N^{\mathsf{th}}$ layer is reconstructed as $\mathsf{vec}\left(\boldsymbol{\widehat{H}}_{\mathsf{c}}\mathit{\left[k\right]}\right)=\boldsymbol{\Psi}\mathsf{vec}\left(\boldsymbol{\widehat{X}}_{N}[k]\right)$. 

\begin{algorithm}
\caption{Parameter Learning of DA-RLAMP Network via Layer-by-Layer Training
Strategy}

\begin{algorithmic}[1]

\Require Complex-valued training set $\left\{ \boldsymbol{Y}_{\mathsf{tra}},\boldsymbol{H}_{\mathsf{tra}}\right\} $

\State Initialization: $\boldsymbol{\beta}_{1}=\text{\ensuremath{\boldsymbol{\Upsilon}}}^{\mathsf{H}}$
and $\boldsymbol{\lambda}_{1}=\left\{ 1\text{,1,1}\right\} $

\State Learn $\left\{ \boldsymbol{\beta}_{1},\boldsymbol{\Theta}\right\} $
to minimize $L_{1}^{\mathsf{L}}$

\State Learn $\boldsymbol{\lambda}_{1}$ with fixed $\left\{ \boldsymbol{\beta}_{1},\boldsymbol{\Theta}\right\} $
to minimize $L_{1}^{\mathsf{NL}}$

\State Refine $\left\{ \boldsymbol{\beta}_{1},\boldsymbol{\Theta},\boldsymbol{\lambda}_{1}\right\} $
to minimize $L_{1}^{\mathsf{NL}}$

\For{$n=2,3\cdots,N$}

\State Initialization: $\boldsymbol{\beta}_{n}=\boldsymbol{\beta}_{n-1}$
and $\boldsymbol{\lambda}_{n}=\boldsymbol{\lambda}_{n-1}$

\State Learn $\boldsymbol{\beta}_{n}$ with fixed $\left\{ \boldsymbol{\Theta},\left\{ \boldsymbol{\beta}_{l},\boldsymbol{\lambda}_{l}\right\} _{l=1}^{n-1}\right\} $
to minimize $L_{n}^{\mathsf{L}}$

\State Refine $\left\{ \boldsymbol{\Theta},\boldsymbol{\beta}_{n},\left\{ \boldsymbol{\beta}_{l},\boldsymbol{\lambda}_{l}\right\} _{l=1}^{n-1}\right\} $
to minimize $L_{n}^{\mathsf{L}}$

\State Learn $\boldsymbol{\lambda}_{n}$ with fixed $\left\{ \boldsymbol{\Theta},\boldsymbol{\beta}_{n},\left\{ \boldsymbol{\beta}_{l},\boldsymbol{\lambda}_{l}\right\} _{l=1}^{n-1}\right\} $
to

\Statex $\hspace{1.5em}$minimize $L_{n}^{\mathsf{NL}}$

\State Refine $\left\{ \boldsymbol{\Theta},\left\{ \boldsymbol{\beta}_{l},\boldsymbol{\lambda}_{l}\right\} _{l=1}^{n}\right\} $
to minimize $L_{n}^{\mathsf{NL}}$

\EndFor

\Ensure Return $\left\{ \boldsymbol{\Theta},\left\{ \boldsymbol{\beta}_{l},\boldsymbol{\lambda}_{l}\right\} _{l=1}^{N}\right\} $

\end{algorithmic}

\label{alg:train DA-RLAMP}
\end{algorithm}
 Based on the structure of DA-RLAMP and the learning strategy in \cite{LAMP},
we propose a novel layer-by-layer training strategy to jointly learn
parameters in the denoising network, attention network and RLAMP network.
Specifically, in the training phase, we obtain the training dataset
$\left\{ \boldsymbol{Y}_{\mathsf{tra}},\boldsymbol{H}_{\mathsf{tra}}\right\} $
according to (\ref{eq:Y_U}) and (\ref{eq:cascaded channel}), where
$\boldsymbol{Y}_{\mathsf{tra}}\in\mathbb{C}^{N_{\mathsf{train}}\times M\times K}$,
$\boldsymbol{H}_{\mathsf{tra}}\in\mathbb{C}^{N_{\mathsf{train}}\times N_{\mathsf{b}}N_{\mathsf{i}}\times K}$
and $N_{\mathsf{train}}$ denote network input, corresponding label,
and the number of training data, respectively. The whole training
phase is divided into a procedure for the first layer learning and
$N-1$ sequential sub-procedures for the rest layers in Algorithm
\ref{alg:train DA-RLAMP}. Therefore, we derive two types of loss
functions:
\begin{align}
 & L_{n}^{\mathsf{L}}\left\{ \boldsymbol{\Theta},\left\{ \boldsymbol{\beta}_{l},\boldsymbol{\lambda}_{l}\right\} _{l=1}^{n}\right\} \nonumber \\
 & =\frac{1}{N_{\mathsf{train}}}\sum_{j=1}^{N_{\mathsf{train}}}\frac{\left\Vert \boldsymbol{\Psi}\boldsymbol{R}_{n}^{j}-\boldsymbol{H}_{\mathsf{tra}}^{j}\right\Vert _{\mathsf{2}}^{2}}{\left\Vert \boldsymbol{H}_{\mathsf{tra}}^{j}\right\Vert _{\mathsf{2}}^{2}}\\
 & =\frac{1}{N_{\mathsf{train}}}\sum_{j=1}^{N_{\mathsf{train}}}\frac{\left\Vert \boldsymbol{\Psi}\mathcal{F}_{n}^{\mathsf{L}}\left(\boldsymbol{Y}_{\mathsf{tra}}^{j};\left\{ \boldsymbol{\Theta},\left\{ \boldsymbol{\beta}_{l},\boldsymbol{\lambda}_{l}\right\} _{l=1}^{n}\right\} \right)-\boldsymbol{H}_{\mathsf{tra}}^{j}\right\Vert _{\mathsf{2}}^{2}}{\left\Vert \boldsymbol{H}_{\mathsf{tra}}\right\Vert _{\mathsf{2}}^{2}},\nonumber 
\end{align}
\begin{align}
 & L_{n}^{\mathsf{NL}}\left\{ \boldsymbol{\Theta},\left\{ \boldsymbol{\beta}_{l},\boldsymbol{\lambda}_{l}\right\} _{l=1}^{n}\right\} \nonumber \\
 & =\frac{1}{N_{\mathsf{train}}}\sum_{j=1}^{N_{\mathsf{train}}}\frac{\left\Vert \widehat{\boldsymbol{H}}_{\mathsf{tra}}^{j}-\boldsymbol{H}_{\mathsf{tra}}^{j}\right\Vert _{\mathsf{2}}^{2}}{\left\Vert \boldsymbol{H}_{\mathsf{tra}}^{j}\right\Vert _{\mathsf{2}}^{2}}\\
 & =\frac{1}{N_{\mathsf{train}}}\sum_{j=1}^{N_{\mathsf{train}}}\frac{\left\Vert \boldsymbol{\Psi}\mathcal{F}_{n}^{\mathsf{NL}}\left(\boldsymbol{Y}_{\mathsf{tra}}^{j};\left\{ \boldsymbol{\Theta},\left\{ \boldsymbol{\beta}_{l},\boldsymbol{\lambda}_{l}\right\} _{l=1}^{n}\right\} \right)-\boldsymbol{H}_{\mathsf{tra}}^{j}\right\Vert _{\mathsf{2}}^{2}}{\left\Vert \boldsymbol{H}_{\mathsf{tra}}\right\Vert _{\mathsf{2}}^{2}},\nonumber 
\end{align}
where $L_{n}^{\mathsf{L}}$ ($L_{n}^{\mathsf{NL}}$) and $\mathcal{F}_{n}^{\mathsf{L}}$
($\mathcal{F}_{n}^{\mathsf{NL}}$) correspond to the linear (nonlinear
shrinkage) loss function and linear (nonlinear) operation of the $n^{\mathsf{th}}$
layer. As shown in Algorithm\ref{alg:train DA-RLAMP}, firstly, we
adopt a backpropagation algorithm to jointly optimize the learnable
parameters $\boldsymbol{\beta}_{1}$ and $\boldsymbol{\Theta}$ by
minimizing $L_{1}^{\mathsf{L}}$ in the first training sub-procedure.
Secondly, the nonlinear shrinkage parameters $\boldsymbol{\lambda}_{1}$
are obtained by nonlinear training processing. At last, all the variables
in the first layer, i.e., $\left\{ \boldsymbol{\beta}_{1},\boldsymbol{\Theta},\boldsymbol{\lambda}_{1}\right\} $
is refined by minimizing $L_{1}^{\mathsf{NL}}$. Similar to the first
layer, the linear transform parameter $\boldsymbol{\beta}_{n}$ and
$\boldsymbol{\lambda}_{n}$ are learned in $n^{\mathsf{th}}$ linear
and nonlinear training phase, respectively. Then all the previous
parameters are optimized jointly. 

\subsection{Mobile DA-RLAMP (MDA-RLAMP) Network for Hybrid IRS Aided Channel
Estimation \label{subsec:denoiser-1}}

\begin{figure}
\vspace{-0.6cm}

\centering\subfloat[]{\centering\includegraphics[height=2in]{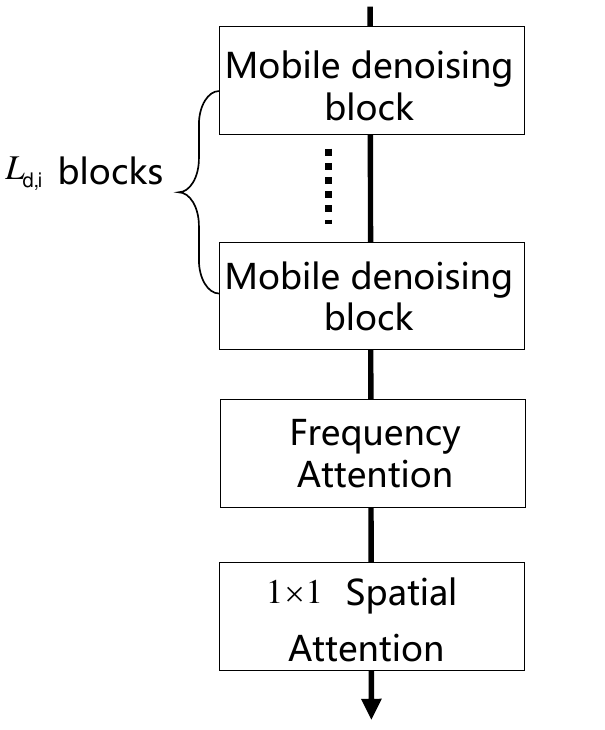}}\subfloat[]{\centering\includegraphics[height=2in]{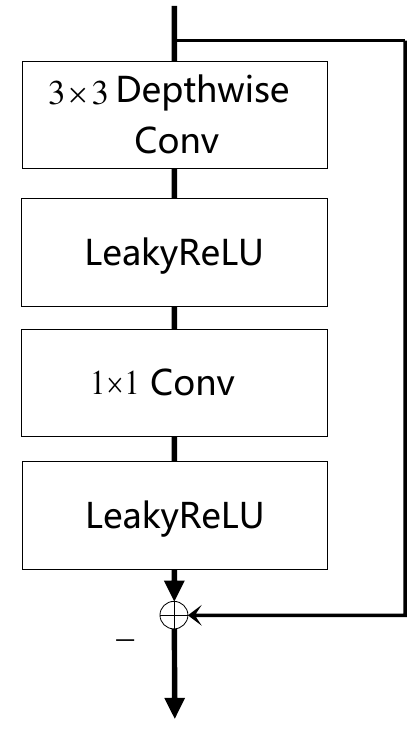}}\caption{(a) Framework of the mobile denoising and attention network (MDA)
in the hybrid IRS aided channel estimator. (b) Overview of the mobile
denoising block.}

\vspace{-0.8cm}

\label{fig:MDA}
\end{figure}

As discussed previously, to overcome these issues and implement hybrid
IRS systems in practice, we develop a more efficient denoising and
attention RLAMP architecture based on mobile architecture. The proposed
mobile DA-RLAMP (MDA-RLAMP) architecture is shown in Fig. \ref{fig:MDA},
which follows the same implementation as that of Algorithm \ref{alg:DA-RLAMP}
except for differences in denoising and attention architecture. Using
the superscript $\mathsf{i}$ for referring to the MDA-RLAMP network,
we explain how to utilize depthwise convolution in \cite{Mobilenet}
instead of the common convolution. 

\emph{1) Mobile Denoising Blocks:} Different from the DnCNN and MobileNet
architecture \cite{Mobilenet}, we design $L_{\mathsf{d,i}}$ mobile
denoising blocks to iteratively learn the residual noise from $\boldsymbol{C}_{\mathsf{a,i}}=\mathcal{F}_{\mathsf{RV}}\left(\boldsymbol{C}_{\mathsf{i}}^{'}\right)$,
where correlation matrix $\boldsymbol{C}_{\mathsf{i}}^{'}\in\mathbb{C}^{\sqrt{G_{\mathsf{i}}}\times\sqrt{G_{\mathsf{i}}}\times K}$
is similar to the $\boldsymbol{C}^{'}$ in (\ref{eq:vec_c}). The
filtering and combining steps of a conventional CNN layer can be split
into two separate layers by utilizing a depthwise separable convolution
and a pointwise convolution in per block. All the blocks in Fig. \ref{fig:MDA}
(a) have an identical structure, which consists of a $3\times3$ depthwise
convolution layer, a $1\times1$ point convolution layer and a subtract
operator, where the depthwise convolution layer with a filter is expressed
as:
\begin{equation}
\boldsymbol{C}_{\mathsf{i},j}^{\mathsf{dc}}[k]=\varphi\left(\mathcal{F}_{\mathsf{D-conv}}\left(\boldsymbol{C}_{\mathsf{i}}[k];\boldsymbol{\phi}_{\mathsf{d},j}\right)\right)\text{,}
\end{equation}
where $\mathcal{F}_{\mathsf{D-conv}}\left(\cdot,\boldsymbol{\phi}_{\mathsf{d},j}\right)$
denotes the depthwise convolution with parameters $\boldsymbol{\phi}_{\mathsf{d},j}$
in the $j^{\mathsf{th}}$ ($j\in\{0\text{,1,}\cdots L_{\mathsf{d,i}}\}$)
mobile denoising block. Note that $\mathcal{F}_{\mathsf{D-conv}}$
adopts a single filter to extract the noise feature for each subcarrier
(input depth). For the last layer, a simple point convolution layer
is used to create a linear combination of the $\boldsymbol{C}_{\mathsf{i},j}^{\mathsf{dc}}[k]$,
\begin{equation}
\boldsymbol{C}_{\mathsf{i},j}^{\mathsf{pc}}[k]=\varphi\left(\mathcal{F}_{\mathsf{\mathsf{conv}}}\left(\boldsymbol{C}_{\mathsf{i},j}^{\mathsf{dc}}[k];\boldsymbol{\phi}_{\mathsf{p},j}\right)\right),
\end{equation}
where $\boldsymbol{\phi}_{\mathsf{p},j}$ denotes the parameters of
the pointwise convolution layer in the $j^{\mathsf{th}}$ block. An
element-wise subtraction connects the input and the output of the
denoising block to obtain the noisy-clean matrix $\boldsymbol{G}_{\mathsf{D,i}}$
gradually. The output of the mobile denoising blocks can be written
as
\begin{equation}
\boldsymbol{G}_{\mathsf{D,i}}[k]=\boldsymbol{C}_{\mathsf{i}}-\sum_{j=1}^{L_{\mathsf{d,i}}}\boldsymbol{D}_{\mathsf{i},j}\left(\boldsymbol{C}_{\mathsf{a,i}};\boldsymbol{\phi}_{j}\right),
\end{equation}
where $\boldsymbol{D}_{\mathsf{i},j}$ and $\boldsymbol{\phi}_{j}=\{\boldsymbol{\phi}_{\mathsf{d},j},\boldsymbol{\phi}_{\mathsf{p},j}\}$
denote the function expression and the parameters of the $j^{\mathsf{th}}$
block, $\Sigma_{j=1}^{L_{\mathsf{d,i}}}\boldsymbol{D}_{\mathsf{i},j}\left(\boldsymbol{C}_{\mathsf{i},j-1}[k],\boldsymbol{\phi}_{j}\right)$
represents the residual noise component. Note that the spatial attention
network in MDA-RLAMP also employs a simple $1\times1$ standard convolution
layer to enhance sparse features.
\begin{algorithm}
\caption{Parameter Learning of MDA-RLAMP Network via Layer-by-Layer Training
Strategy}

\begin{algorithmic}[1]

\Require Complex-valued training set $\left\{ \boldsymbol{Y}_{\mathsf{\mathsf{i},tra}},\boldsymbol{F}_{\mathsf{tra}}\right\} $

\State Initialization: $\boldsymbol{\beta}_{\mathsf{i}}=\text{\ensuremath{\boldsymbol{\Upsilon}}}_{\mathsf{i}}^{\mathsf{H}}$
and $\boldsymbol{\lambda}_{\mathsf{i},1}=\left\{ 1\text{,1,1}\right\} $

\State Learn $\left\{ \boldsymbol{\beta}_{\mathsf{i}},\boldsymbol{\boldsymbol{\phi}}\right\} $
to minimize $L_{1}^{\mathsf{L,i}}$

\For{ $n=2,3\cdots,N$}

\State Initialization: $\boldsymbol{\lambda}_{\mathsf{i},n}=\boldsymbol{\lambda}_{\mathsf{i},n-1}$

\State Learn $\boldsymbol{\lambda}_{\mathsf{i},n}$ with fixed $\left\{ \boldsymbol{\boldsymbol{\phi}},\boldsymbol{\beta}_{\mathsf{i}},\left\{ \boldsymbol{\lambda}_{\mathsf{i},l}\right\} _{l=1}^{n-1}\right\} $
to minimize $L_{n}^{\mathsf{L,i}}$

\State Re-learn $\left\{ \boldsymbol{\boldsymbol{\phi}},\boldsymbol{\beta}_{\mathsf{i}},\left\{ \boldsymbol{\lambda}_{\mathsf{i},l}\right\} _{l=1}^{n}\right\} $
to minimize $L_{n}^{\mathsf{L,i}}$

\EndFor

\Ensure Trainable variables$\left\{ \boldsymbol{\boldsymbol{\phi}},\boldsymbol{\beta}_{\mathsf{i}},\left\{ \boldsymbol{\lambda}_{\mathsf{i},l}\right\} _{l=1}^{N}\right\} $

\end{algorithmic}

\label{alg:train MDA-RLAMP}
\end{algorithm}

\emph{2) Parameters Learning Strategy: }Considering the number of
parameters and training complexity of the hybrid IRS, the parameter
$\boldsymbol{\beta}_{\mathsf{i}}$ is fixed for all layers in the
MDA-RLAMP network and $\text{\ensuremath{\boldsymbol{\Upsilon}}}_{\mathsf{i}}=\boldsymbol{\Phi}_{\mathsf{i}}\boldsymbol{\Psi}_{\mathsf{i}}$.
We simplify the layer-by-layer training strategy, which is shown in
Algorithm \ref{alg:train MDA-RLAMP}. In specific, $\left\{ \boldsymbol{\beta}_{\mathsf{i}},\boldsymbol{\boldsymbol{\phi}}\right\} $
($\boldsymbol{\boldsymbol{\phi}}=\left[\boldsymbol{\phi}_{1},\cdots\boldsymbol{\phi}_{2},\cdots\boldsymbol{\phi}_{j}\right]$)
is firstly trained in line 2 by minimizing the loss function $L_{1}^{\mathsf{L,i}}$
of the first layer and updated from the training epoch of $n^{\mathsf{th}}$
($n=1,2\cdots,N$) layers for refinement. To minimize $L_{n}^{\mathsf{L,i}}$,
the shrinkage parameter $\boldsymbol{\lambda}_{\mathsf{i},n}$ of
the $n^{\mathsf{th}}$ layer is learned individually in line 5 and
jointly updated with $\left\{ \boldsymbol{\boldsymbol{\phi}},\boldsymbol{\beta}_{\mathsf{i}},\left\{ \boldsymbol{\lambda}_{\mathsf{i},l}\right\} _{l=1}^{n-1}\right\} $
in line 6. We define the loss function of the $n^{\mathsf{th}}$ layer
as follows 
\begin{align}
 & L_{n}^{\mathsf{L,i}}\left\{ \boldsymbol{\boldsymbol{\phi}},\boldsymbol{\beta}_{\mathsf{i}},\left\{ \boldsymbol{\lambda}_{\mathsf{i},l}\right\} _{l=1}^{n}\right\} \nonumber \\
 & =\frac{1}{N_{\mathsf{i\text{,}train}}}\sum_{j=1}^{N_{\mathsf{i,train}}}\frac{\left\Vert \boldsymbol{\widehat{F}}_{n}^{j}-\boldsymbol{F}_{\mathsf{tra}}^{j}\right\Vert _{\mathsf{2}}^{2}}{\left\Vert \boldsymbol{F}_{\mathsf{tra}}^{j}\right\Vert _{\mathsf{F}}^{2}}\\
 & =\frac{1}{N_{\mathsf{i\text{,}train}}}\sum_{j=1}^{N_{\mathsf{i,train}}}\frac{\left\Vert \boldsymbol{\Psi}_{\mathsf{i}}\mathcal{F}_{n}^{\mathsf{i}}\left(\boldsymbol{Y}_{\mathsf{\mathsf{i,tra}}}^{j},\left\{ \boldsymbol{\boldsymbol{\phi}},\boldsymbol{\beta}_{\mathsf{i}},\left\{ \boldsymbol{\lambda}_{\mathsf{i},l}\right\} _{l=1}^{n}\right\} \right)-\boldsymbol{F}_{\mathsf{tra}}^{j}\right\Vert _{\mathsf{2}}^{2}}{\left\Vert \boldsymbol{F}_{\mathsf{tra}}^{j}\right\Vert _{\mathsf{2}}^{2}},\nonumber 
\end{align}
where $\mathcal{F}_{n}^{\mathsf{i}}\left(\cdot\text{,}\cdot\right)$
denotes the function expression of the $n^{\mathsf{th}}$ layer and
$\boldsymbol{Y}_{\mathsf{i,tra}}\in\mathbb{C}^{N_{\mathsf{i\text{,}train}}\times M_{\mathsf{i}}\times K},\boldsymbol{F}_{\mathsf{tra}}\in\mathbb{C}^{N_{\mathsf{i\text{,}train}}\times N_{\mathsf{i}}\times K}$
denotes $N_{\mathsf{i\text{,}train}}$ training data pairs. Compared
with the strategy in Algorithm \ref{alg:train DA-RLAMP}, the learning
strategy of MDA-RLAMP simplifies the training process, which accelerates
the convergence and enables a rapid deployment on hybrid IRS. 

\section{Numerical Results }

\begin{figure}
\vspace{-0.6cm}

\centering\includegraphics[width=3.5in]{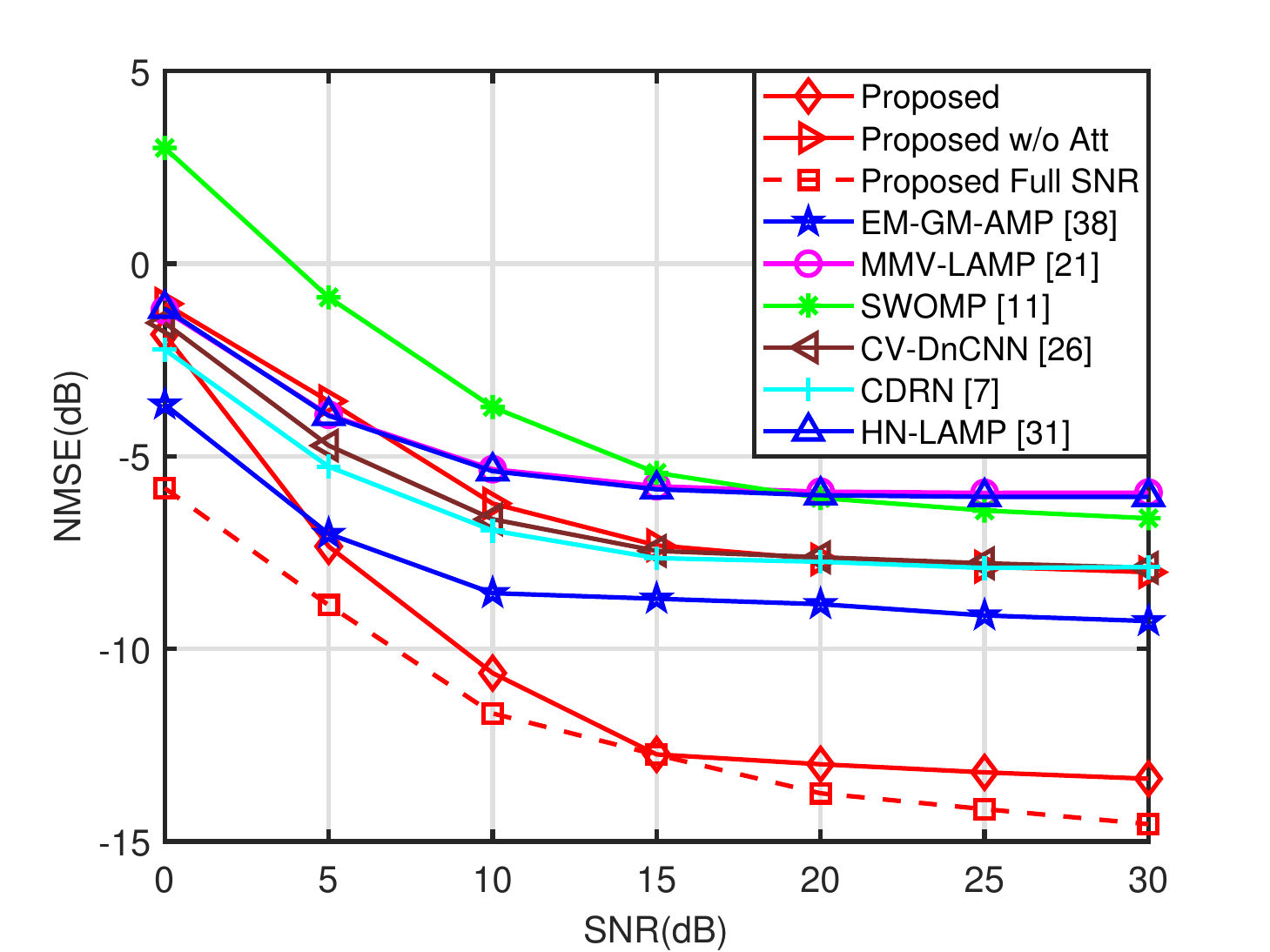}\caption{Evolution of the NMSE versus SNR for the different estimation algorithms.
The proposed approach is pretrained with SNR = 15 dB, $T=32$.}

\vspace{-0.6cm}

\label{fig:NMSE_ASE}
\end{figure}
 In this section, we evaluate the performance and computational complexity
of proposed channel estimation algorithms with other frequency-domain
estimation schemes. The simulations are performed based on the widely
used Saleh-Valenzuela channel model. Then, the passive IRS and hybrid
IRS channel estimation results are provided through extensive Monte
Carlo simulations.

\subsection{Simulation Settings}

In our simulations, a BS is equipped with $N_{\mathsf{b}}=16$ antennas
and an IRS with $N_{\mathsf{i}}=64$ phase shifters. Other parameters
are shown in Table \ref{table:sim_paras}. For a fair comparison,
the simulation parameters used in our work are similar to \cite{wang2020compressed,GAN_CE,CVDNCNN};
The dataset is divided into the training dataset and test dataset
randomly. In addition, the experiments are simulated on a computer
with an Nvidia GeForce GTX 3090 GPU and an AMD 5950X CPU. 
\begin{table}[htbp]
\vspace{-0.5cm}

\caption{Simulation parameters of Passive IRS scenario}

\renewcommand\arraystretch{0.6}

\centering%
\begin{tabular}{cc}
\hline 
Parameter & Value\tabularnewline
\hline 
Total size of training dataset & 20000\tabularnewline
Total size of test dataset & 2000\tabularnewline
Total number of subcarriers $K$ & 16\tabularnewline
Total number of UEs $U$ & 4\tabularnewline
Operating frequency & 100GHz\tabularnewline
Max multipath delay & 100ns\tabularnewline
Channel path $L_{\mathsf{p,f}}$ ($L_{\mathsf{p,g}}$) & 5\tabularnewline
Distribution of AoAs & $u(0\text{,}\pi)$\tabularnewline
Oversampling ratio of BS (IRS) dictionary $G_{\mathsf{b}}(G_{\mathsf{i}})$ & $2N_{\mathsf{b}}(4N_{\mathsf{i}})$\tabularnewline
\hline 
\end{tabular}

\vspace{-0.5cm}

\label{table:sim_paras}
\end{table}

We compare the proposed algorithms with the following benchmark algorithms.
The structures of all DL-based baselines are carefully simulated by
cross-validation.
\begin{itemize}
\item Conventional estimation methods: The SWOMP method \cite{SOMP18} and
the EM-GM-GAMP algorithm \cite{EM-GM-AMP} are designed for the estimation
problem. The maximum number of iterations of both algorithms is set
to 100, and the channel in (\ref{eq:problem}) exhibits the same sparse
structure for all subcarriers. 
\item Model driven methods: The MMV-LAMP structure with thresholding shrinkage
function \cite{MMVLAMP} was compared under the same simulation conditions
while the number of layers was set to 6. The HN-LAMP \cite{HN-LAMP}
with the hypernetwork consists of two layers containing 128 and 1
neurons, respectively. Both methods are well-trained under SNR = 15
dB. Note that the training strategies in passive and hybrid IRS aided
systems are shown in Algorithm \ref{alg:train DA-RLAMP} and Algorithm
\ref{alg:train MDA-RLAMP}, respectively. The learning rate is initialized
as 0.001.
\item Data driven methods: The frequency-domain channel is first reconstructed
by the SWOMP algorithm. Then the complex network CV-DnCNN \cite{CVDNCNN}
and CDRN \cite{CDRN} are used to learn both the features and noise.
The training strategies and network parameters are provided in \cite{CVDNCNN}
and \cite{CDRN}, respectively.
\item Proposed: The structure of the proposed DA-RLAMP is used to estimate
channels from the received pilots. The proposed network is composed
of $N=6$ layers, and the DA network with $L_{\mathsf{d}}=3$ convolutional
layers. The proposed networks are trained for 10000 epochs in each
iteration. Meanwhile, the Adam algorithm \cite{adam} is used as the
weight optimizer, and the learning rate is initialized as 0.001. 
\item Proposed w/o Att: The same structure and training strategies but with
all the attention modules removed.
\item Proposed Full SNR: The proposed scheme is trained and tested under
the same SNR.
\end{itemize}
The NMSE and ergodic spectral efficiency metric are chosen for the
quantitative evaluation of estimation algorithms, where the NMSE is
defined as 
\begin{equation}
\textrm{NMSE}\left(\boldsymbol{H},\widehat{\boldsymbol{H}}\right)=10\textrm{lg}\left[\mathbb{E}\left(\frac{\left\Vert \mathsf{vec}\left(\boldsymbol{H}-\widehat{\boldsymbol{H}}\right)\right\Vert _{2}^{2}}{\left\Vert \mathsf{vec}\left(\boldsymbol{H}\right)\right\Vert _{2}^{2}}\right)\right],
\end{equation}
where $\widehat{\boldsymbol{H}}$ and $\boldsymbol{H}$ denote the
output of channel estimation and the true channel, respectively. For
simplicity, the ergodic spectral efficiency can be expressed as \cite{Tensor_IRS}
\begin{equation}
E[k]=\textrm{log}_{2}\textrm{det}\left(1+\frac{\left|\left(\boldsymbol{G}[k]\mathrm{diag}\left(\boldsymbol{f}[k]\right)\boldsymbol{r}\right)\boldsymbol{W}_{\mathsf{b}}\right|^{2}}{\sigma^{2}}\right),
\end{equation}
where $E[k]$ denotes ergodic spectral efficiency of $k^{\mathsf{th}}$
subcarrier and $\boldsymbol{W}_{\mathsf{b}}$ donates the precoding
matrix at BS side, while the IRS reflection vector $\boldsymbol{r}$
is generated from the IRS controller. The IRS aided system adopts
the hybrid precoding algorithm in \cite{HBF} to jointly optimize
the $\boldsymbol{r}$ and $\boldsymbol{W}_{\mathsf{b}}$. 

\subsection{Passive IRS Scenario }

\begin{figure*}
\vspace{-0.8cm}

\centering\subfloat[]{\centering\includegraphics[width=2.3in]{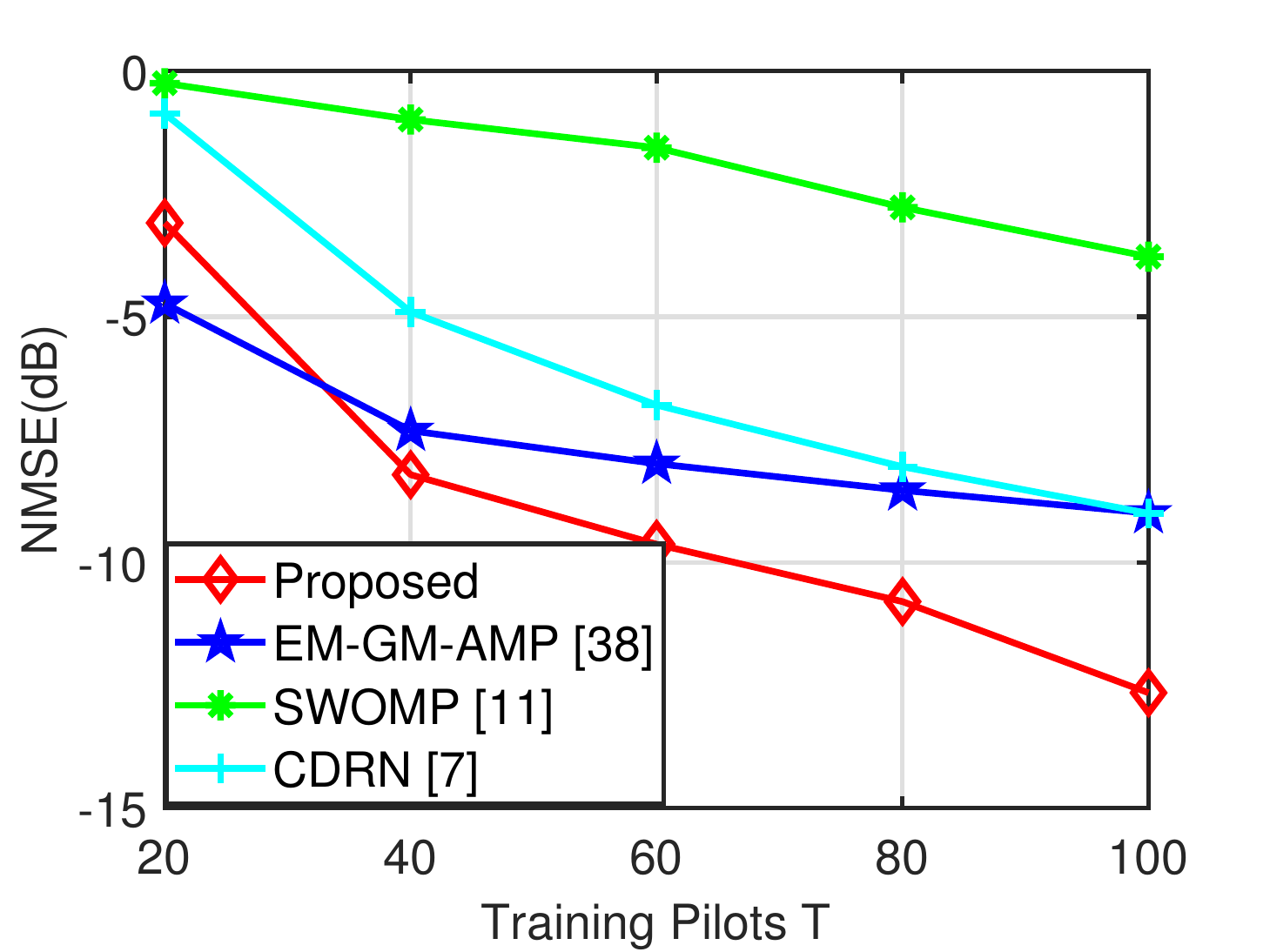}}\subfloat[]{\centering\includegraphics[width=2.3in]{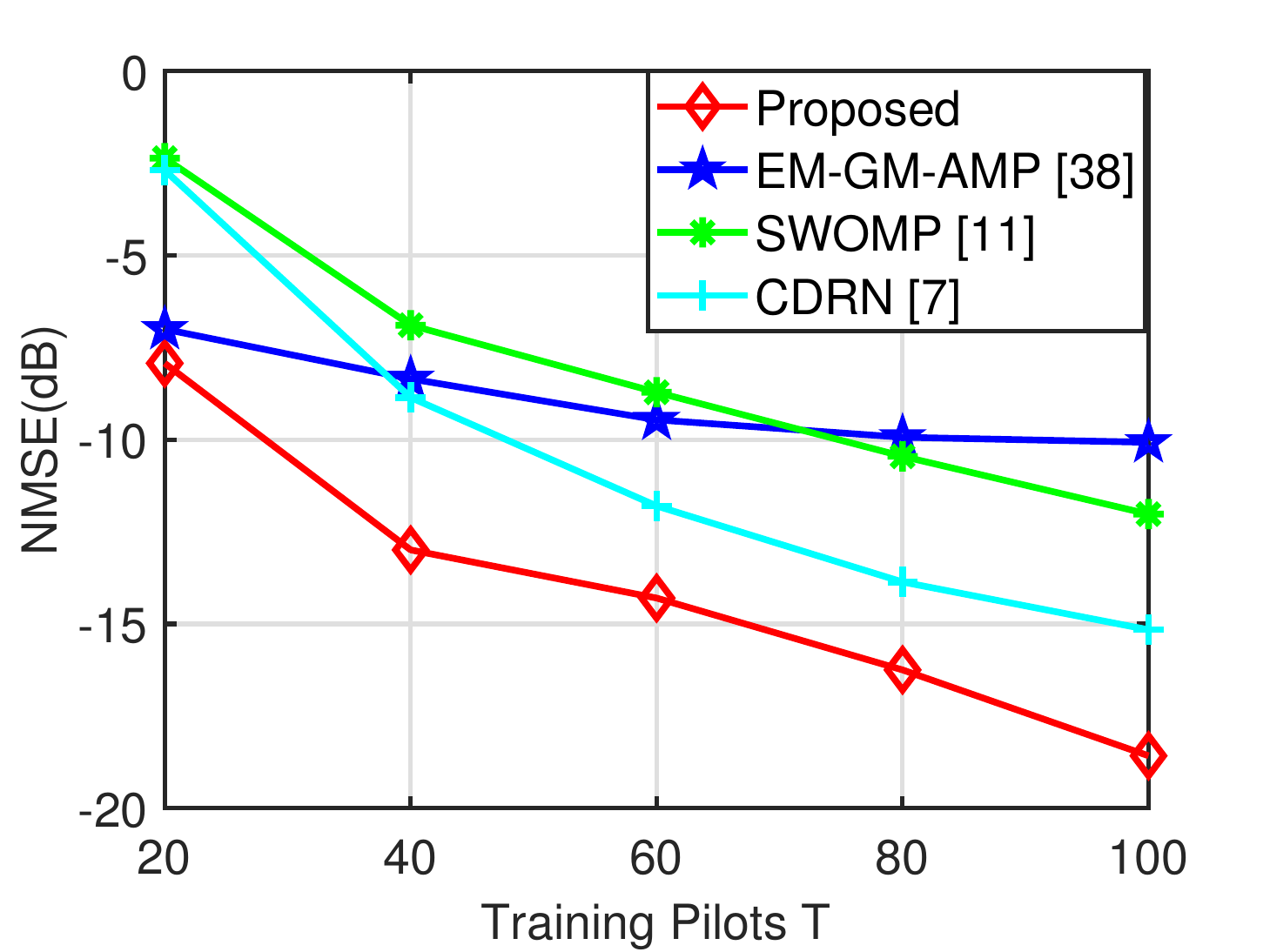}}\subfloat[]{\centering\includegraphics[width=2.3in]{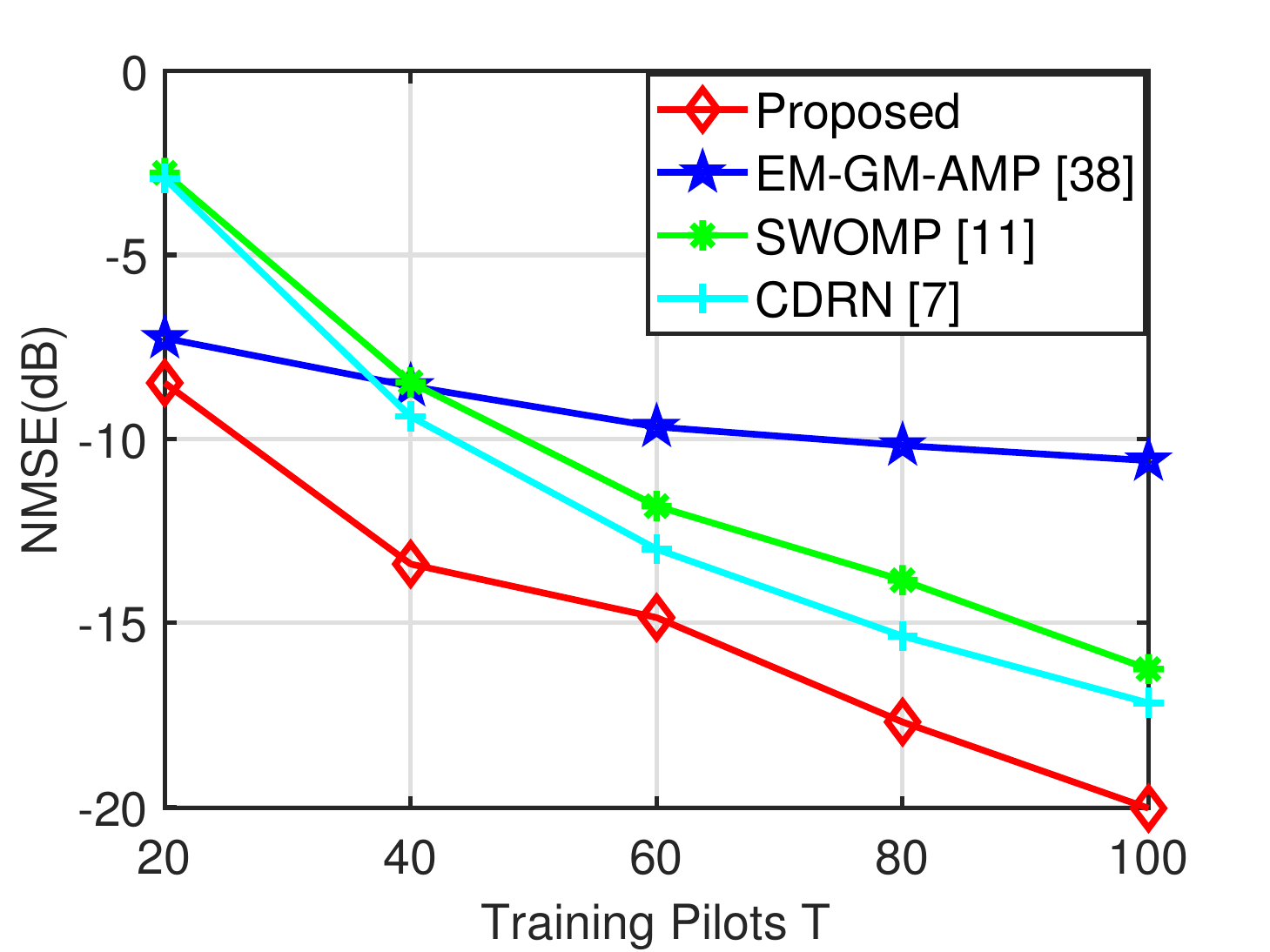}}

\vspace{-0.3cm}

\centering\subfloat[]{\centering\includegraphics[width=2.3in]{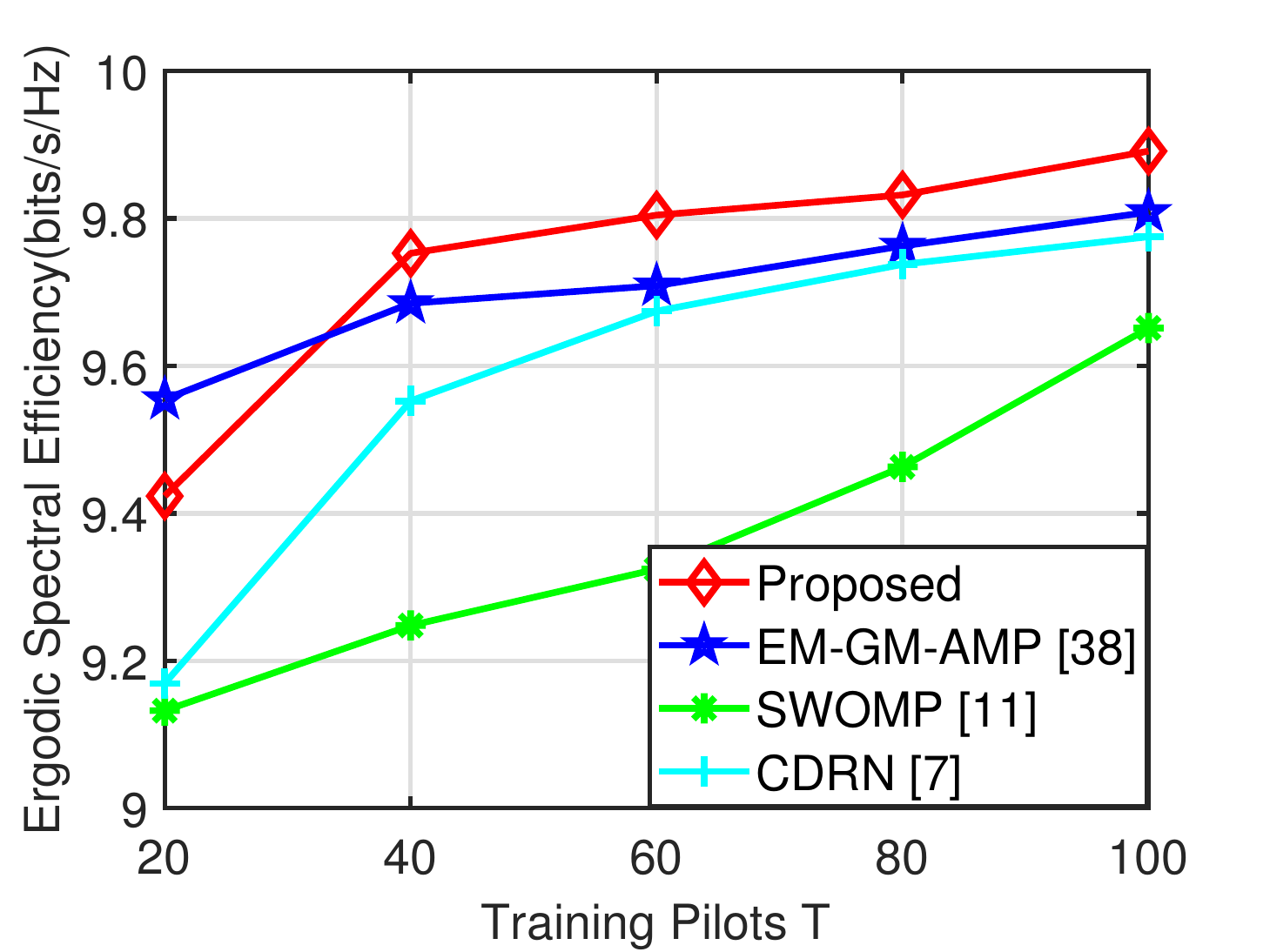}}\subfloat[]{\centering\includegraphics[width=2.3in]{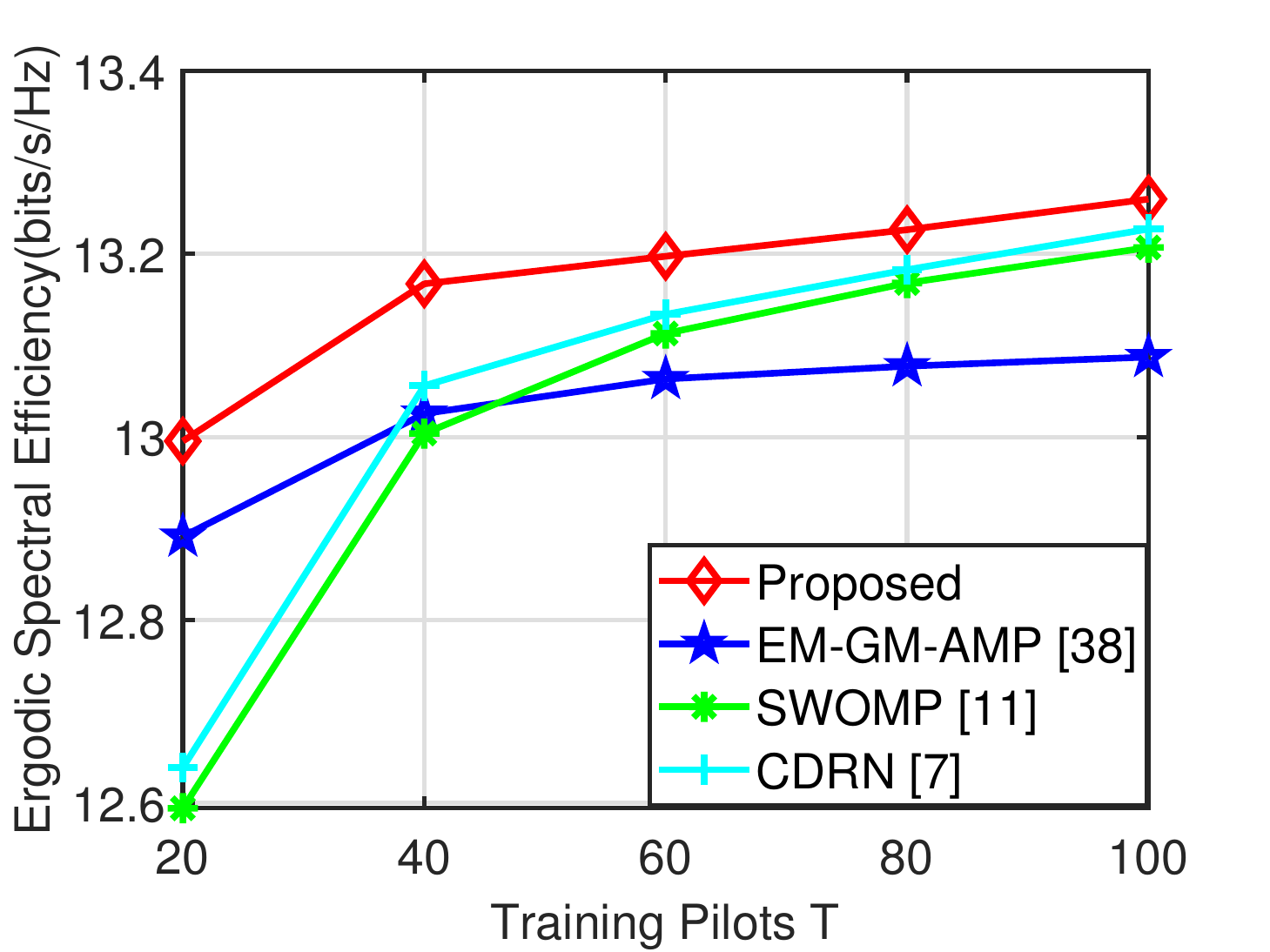}}\subfloat[]{\centering\includegraphics[width=2.3in]{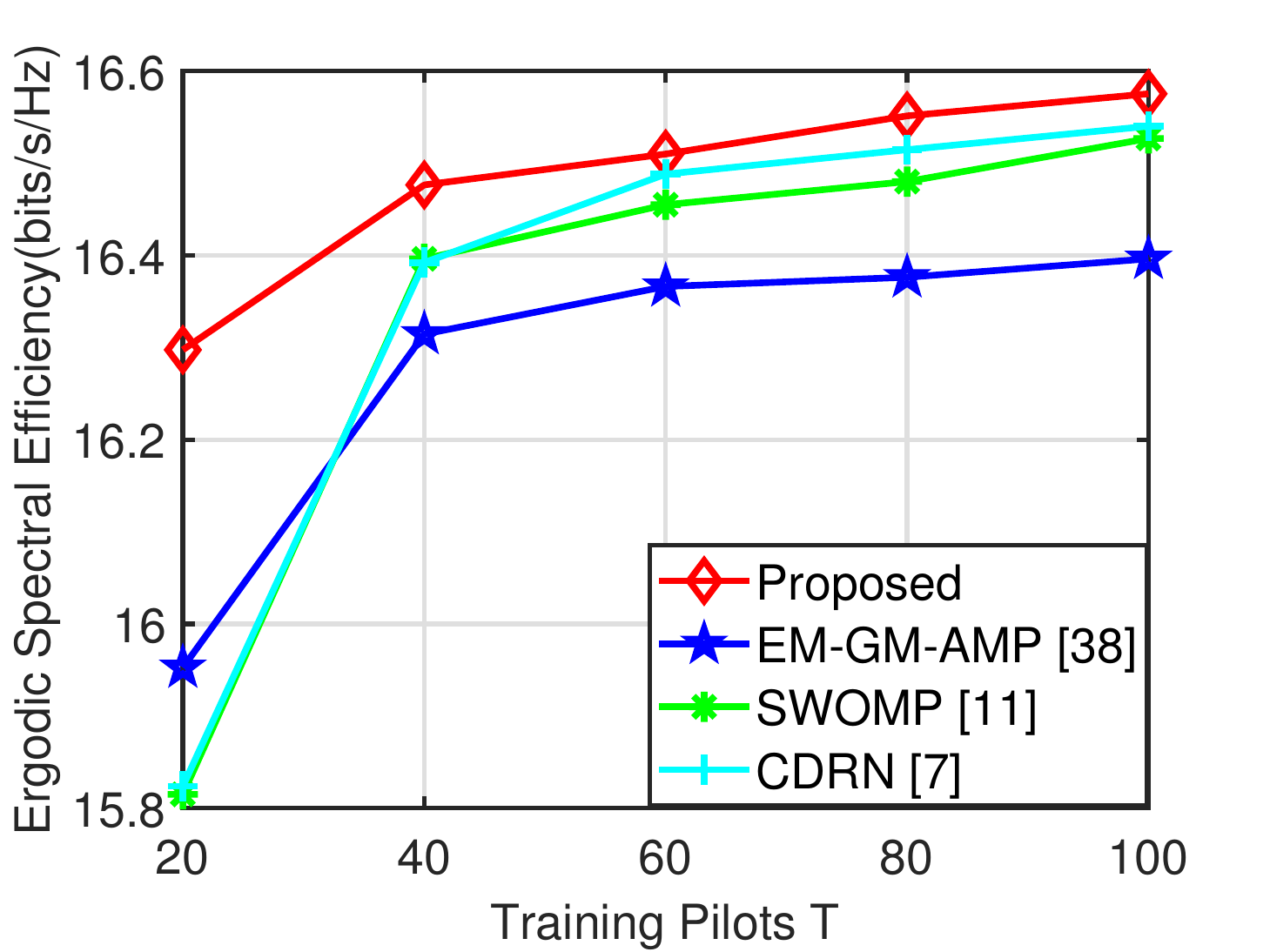}
}

\caption{Comparison of evolution of the NMSE and ergodic spectral efficiency
versus number of training pilots $T$ at different SNRs. The SNR is
set to 0 dB (a) and (d), 15 dB (b) and (e) and 25 dB (c) and (f).
The curves in the first row compare of the NMSE performance for the
different channel estimation algorithms, while the ones in the second
row compare the ergodic spectral efficiency performance.}

\vspace{-0.8cm}

\label{fig:NMSE_ASE_slot}
\end{figure*}
 The results of NMSE performance under different SNRs are depicted
in Fig. \ref{fig:NMSE_ASE} for a practical SNR range of 0 dB to 30
dB and training pilots $T=32$. It can be observed that the proposed
DA-RLAMP trained with a specific SNR is capable of outperforming NMSE
values than conventional, data driven and model driven approaches,
even with plenty of labeled data and powerful architectures. The EM-GM-GAMP
and CDRN algorithms are outperforming the proposed methods for SNRs
below 5 dB. The reason is that the proposed method is trained under
the specific SNR value. By contrast, the NMSE performance difference
between the proposed and other algorithms is noticeable when the SNR
range of 5 dB to 30 dB. Precisely, the DA-RLAMP trained under 15 dB
obviously delivers enjoys lower estimation errors than that of GAMP
and other deep learning networks by -2 dB with SNR = 5 dB. In addition,
the proposed algorithm achieves lower NMSE values (-13 dB) at high
SNR values, such as SNR = 20 dB, while other algorithms with higher
resolution grid sizes at SNR = 20 dB achieve NMSEs between about -6
dB and -9 dB. From the curves shown in Fig. \ref{fig:NMSE_ASE}, one
can observe significant improvement of the hybrid driven approach
thanks to the AMP-based structure and attention mechanism, which exploits
more frequency and spatial features of channels from higher angular
resolution redundant dictionary matrices.

To further investigate the percentage of the pilot overhead, Fig.
\ref{fig:NMSE_ASE_slot} shows the average NMSE and ergodic spectral
efficiency versus the number of training pilots $T$. The improvement
in NMSE and ergodic spectral efficiency performance occurs thanks
to a larger number of training pilots, resulting in smaller estimation
errors via CS. Among the state-of-art estimators, the proposed hybrid
driven approach always outperforms the others for each training pilot.
On the one hand, the SWOMP method performs the worst, which comes
from the fact that SWOMP neglects sparsity coefficients and the error
accumulation of the $K$ parallels estimator \cite{OMP_CE_17}. On
the other hand, though the CDRN mitigates the residual noise and recovers
the channel, the estimation performance is affected by the initial
coarse estimated value, which is the input of the network. More importantly,
we can observe that the proposed scheme can reduce the pilot overhead
even up to $60\%$, $50\%$, and $40\%$ with only around 1 dB loss
in NMSE when SNR is 5 dB, 15 dB and 25 dB, respectively. Therefore,
the proposed algorithm can reliably reconstruct the cascade IRS channel
with less training overhead in wideband IRS aided systems. 

\subsection{Hybrid IRS Scenario }

\begin{figure*}
\vspace{-0.6cm}

\centering\subfloat[]{\centering\includegraphics[width=3.5in]{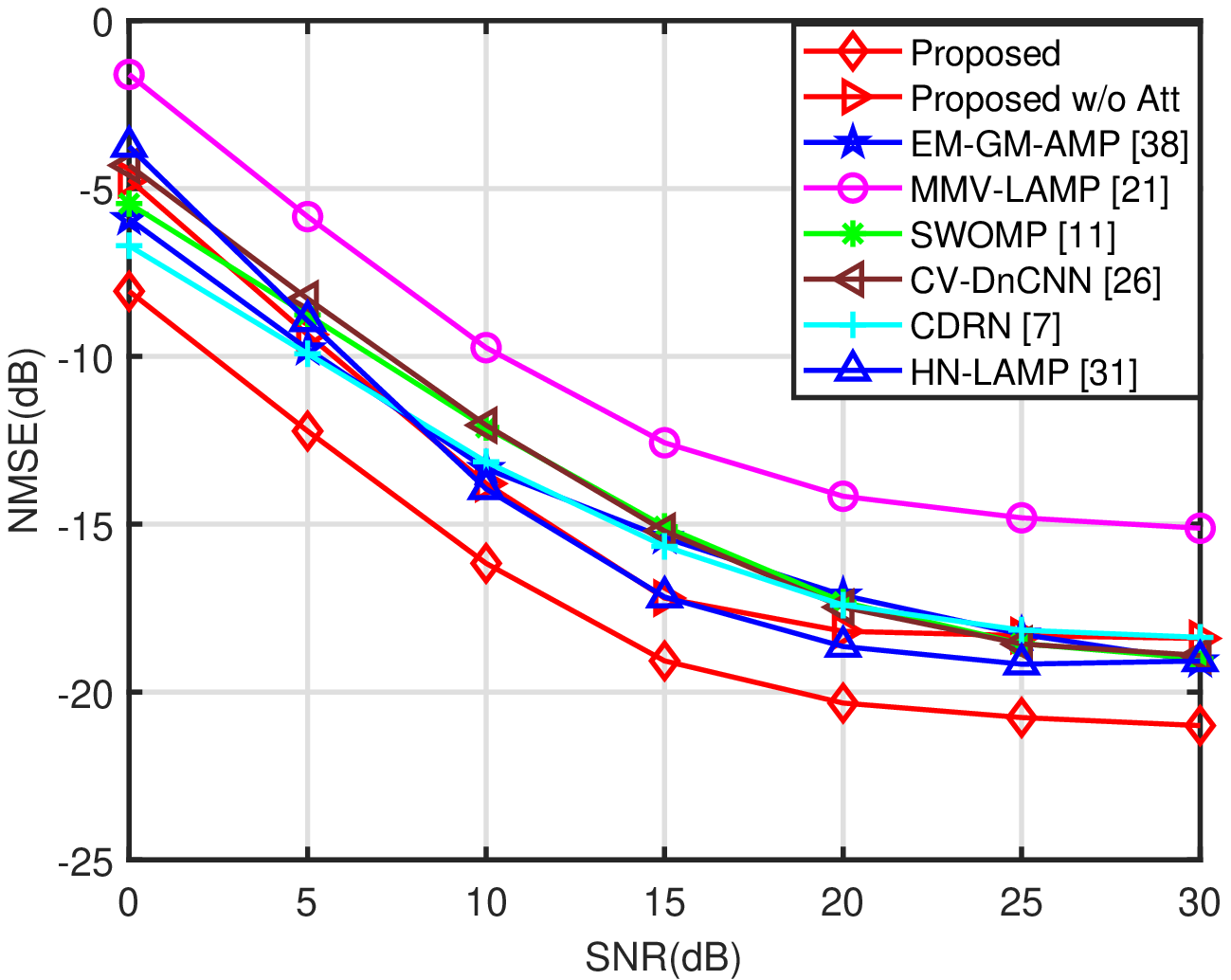}}
\subfloat[]{\centering\includegraphics[width=3.5in]{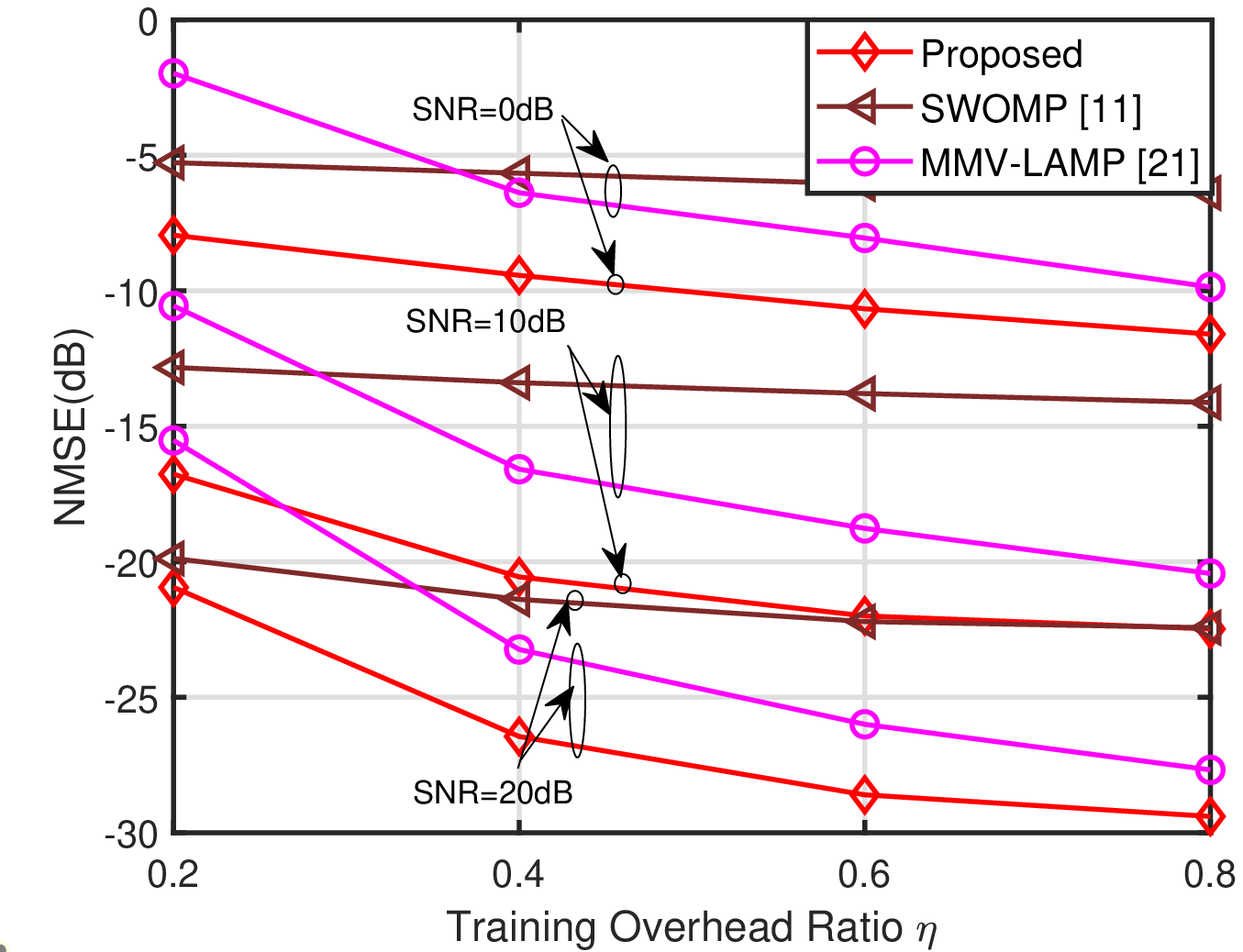}}

\caption{(a) Comparison of the NMSE versus SNR for hybrid IRS architecture
at $T_{\mathsf{i}}=32$, $K=16$ and $G_{\mathsf{i}}=1024$. (b) Comparison
of the NMSE versus training overhead at different SNR for the different
estimation algorithms, where $\eta$ is defined as the ratio of the
number of $T_{\mathsf{i}}$ to the $N_{\mathsf{i}}$.}

\vspace{-0.8cm}

\label{fig:hybrid_NMSE}
\end{figure*}

In this subsection, we evaluate the NMSE values of the proposed scheme
and other solutions for hybrid IRS systems. In Fig. \ref{fig:hybrid_NMSE},
we compare the NMSE performance with the existing channel estimation
methods. We assume that $N_{\mathsf{i}}=256$, $G_{\mathsf{i}}=1024$
and $T_{\mathsf{i}}=32$ active channel sensors are randomly distributed
over the IRS. The remaining parameters in the hybrid scenario are
the same as in Fig. \ref{fig:NMSE_ASE}. Our proposed MDA-RLAMP always
outperforms the six baselines. Concretely, given NMSE $\thickapprox-15$
dB, the proposed MDA-RLAMP method achieves an SNR gain of 4 dB and
6 dB compared with the HN-LAMP and CDRN, respectively. As can be observed,
the NMSEs of MDA-RLAMP decrease with the aid of the attention mechanism.
This demonstrates the competitive advantage of the proposed hybrid
driven approach in exploiting effective frequency and spatial features
for improving recovery accuracy. We further investigate the training
overhead of the proposed scheme as a function of the number of training
pilots $T_{\mathsf{i}}$ in Fig. \ref{fig:hybrid_NMSE} (b). We use
the SWOMP and MMV-LAMP algorithms as benchmarks with SNR $=\left\{ 0,10,20\right\} $
dB. There is a clear performance gain of the schemes indicated in
the curves. Specifically, MDA-RLAMP is the algorithm providing the
best performance with various SNRs, and the MMV-LAMP is sensitive
to the number of pilots. As we can see, the proposed scheme can reduce
the pilot overhead by at least $75\%$ and $50\%$ more than SWOMP
and MMV-LAMP while achieving the same or even better channel estimation
performance. 

\subsection{Comparison  of the Iterations and Computational Complexity}

The above experiments indicate that the hybrid driven methods perform
well in CS. In the following, we further simulate the NMSE performance
against the number of iterations to show the convergence of the proposed
hybrid driven scheme. We compare two proposed schemes with MMV-LAMP
and HN-LAMP, and the curve denoted by \textquotedblleft Proposed w/o
Res\textquotedblright{} means that the proposed schemes without residual
learning mechanism.
\begin{figure}
\vspace{-0.8cm}

\centering\includegraphics[height=2.5in]{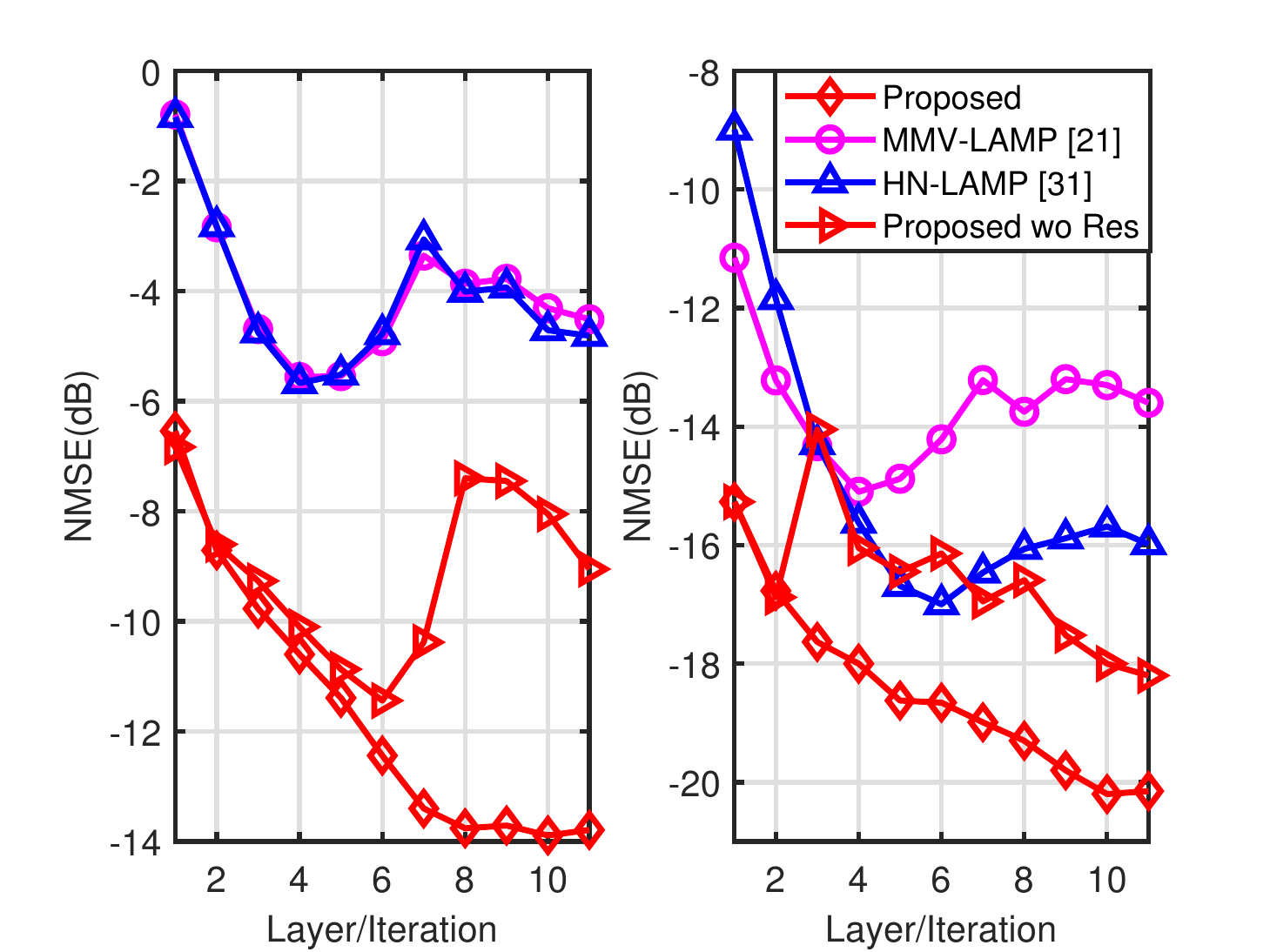}\caption{NMSE versus iterations times under SNR = 15 dB. The curves on the
left consider the passive IRS scenario with the same parameters in
Fig. \ref{fig:NMSE_ASE}. The curves on the right consider the hybrid
IRS scenario with the same parameters in Fig. \ref{fig:hybrid_NMSE}.}

\vspace{-0.8cm}

\label{fig:iteration_fig}
\end{figure}
\begin{table*}
\caption{Computational Complexity and Computation Time of Different Estimation
Schemes}

\renewcommand\arraystretch{0.6}

\centering %
\begin{tabular}{cccc}
\hline 
\multirow{2}{*}{Schemes} & \multirow{2}{*}{Computational Complexity} & Passive IRS & Hybrid IRS\tabularnewline
 &  & Execution Time (ms) & Execution Time (ms)\tabularnewline
\hline 
SWOMP \cite{SOMP18} & $KGT\hat{L}$ & / & /\tabularnewline
EM-GM-GAMP \cite{EM-GM-AMP} & $NKGT$ & / & /\tabularnewline
MMV-LAMP \cite{MMVLAMP} & $NKGT$ & 31.18 & 13.92\tabularnewline
CV-DnCNN \cite{CVDNCNN} & $KGT\hat{L}+9N_{\mathsf{Dn}}N_{\mathsf{i}}N_{\mathsf{b}}k^{2}$ & 1131.68 & 34.84\tabularnewline
CDRN \cite{CDRN} & $KGT\hat{L}+9N_{\mathsf{CD}}D_{\mathsf{CD}}N_{\mathsf{i}}N_{\mathsf{b}}k^{2}$ & 1131.27 & 34.88\tabularnewline
HN-LAMP \cite{HN-LAMP} & $NKGT+128NTK$ & 36.92 & 35.46\tabularnewline
DA-RLAMP & $NKGT+$$9G\left(\left(L_{\mathsf{d}}+1\right)k^{2}+2\right)+6k^{2}$ & 37.16 & /\tabularnewline
MDA-RLAMP & $NKGT+$$G\left(L_{\mathsf{d}}\left(9k+k^{2}\right)+k^{2}+2\right)+6k^{2}$ & / & 17.94\tabularnewline
\hline 
\end{tabular}

\label{table:Alg_name}
\noindent \begin{raggedright}
Note: The learning methods are performed in Python 3.8 and Tensorflow
2.4 environment, while the conventional methods are executed in MATLAB.
Therefore, the execution time of SWOMP and EM-GM-AMP is omitted.
\par\end{raggedright}
\vspace{-0.8cm}
\end{table*}

Fig. \ref{fig:iteration_fig} shows the curves of NMSE versus $N$
for different schemes. It can be observed that the estimation error
of the proposed algorithms without residual learning is undulating
as the number of iterations increases. In specific, the NMSE performance
of DA-LAMP deteriorates when the number of iterations exceeds 6, and
the MDA-LAMP method has a lower convergence rate during the training
period. This is because the vanishing gradients problem hampers convergence
as the number of iterations and parameters increase \cite{ResNet}.
In constant, the proposed algorithms provide faster convergences at
the early stage. Specifically, the DA-RLAMP (MDA-RLAMP) performs about
5 dB (2 dB) better than the network without residual learning and
converges within 8 (10) layers. This phenomenon derives from the fact
that the effectiveness of residual learning in complex networks, where
each layer updates network parameters based on the result and information
from the previous layers.

Besides, we outline the computational complexity of the proposed algorithms
in online deployment\footnote{The complexity of the offline training stage becomes negligible thanks
to the generalization capability of neural networks \cite{GAN_CE}.}. The computational cost of $L$ convolutional layers and $L$ depthwise
convolution layers can be respectively expressed as \cite{Mobilenet}
\begin{equation}
\mathcal{O}\left(\sum_{j=1}^{L}\left(D_{\mathsf{h}}^{j}D_{\mathsf{w}}^{j}s_{\mathsf{k},j}^{2}n_{j-1}n_{j}\right)\right),
\end{equation}
\begin{equation}
\mathcal{O}\left(\sum_{j=1}^{L}\left(D_{\mathsf{h}}^{j}D_{\mathsf{w}}^{j}s_{\mathsf{k},j}^{2}n_{j-1}\right)\right),
\end{equation}
where the $j^{\mathsf{th}}$ convolution layer takes input tensor
with size $D_{\mathsf{h}}^{j}\times D_{\mathsf{w}}^{j}\times n_{j-1}$
and uses kernel with size $s_{\mathsf{k},j}\times s_{\mathsf{k},j}\times n_{j-1}\times n_{j}$.
The complexity of the DA-RLAMP in Algorithm \ref{alg:DA-RLAMP} mainly
stems include: i) The LAMP architecture, i.e., $\mathcal{O}(NKGT)$
for $K$ subcarriers; ii) $L_{\mathsf{d}}$ convolutional layers in
DnCNN with computational complexity $\mathcal{O}\left(\sum_{j=1}^{L_{\mathsf{d}}}\left(Gs_{\mathsf{k},j}^{2}n_{j-1}n_{j}\right)\right)$;
iii) $L_{\mathsf{s}}$ convolutional layers in spatial attention network
with computational complexity $\mathcal{O}\left(\sum_{j=1}^{L_{\mathsf{s}}}\left(Gs_{\mathsf{k},j}^{2}n_{j-1}n_{j}\right)\right)$;
iv) $L_{\mathsf{f}}$ layers in frequency attention network with computational
complexity $\mathcal{O}\left(\sum_{j=1}^{L_{\mathsf{f}}}\left(m_{j-1}m_{j}\right)\right)$,
where $m_{j}$ is the dimensions of the $j^{\mathsf{th}}$ layer output.
As for the SWOMP algorithm, the complexity is $KGT$ for each iteration,
and the algorithm will be repeated for a total of $\hat{L}$ iterations,
where $\hat{L}$ denotes sufficient paths \cite{HAD_CE_TWC22}. While
the computational cost and execution time of the CV-DnCNN and CDRN
come from the traditional channel estimation algorithms and deep learning
networks. As described above, the computational complexity and time
complexity of the proposed and other schemes are summarized in Table
\ref{table:Alg_name}. Thanks to the parallelization of a graphics
processing unit, the execution time in online estimation can be greatly
reduced. By contrast, the DA-RLAMP, MMV-LAMP and HN-LAMP have similar
calculation times in the passive IRS scenario, while DA-RLAMP and
MMV-LAMP consume less time than other schemes in the hybrid IRS scenario.

\section{Conclusions }

In this paper, we have proposed two hybrid driven networks to address
the channel estimation problems for IRS aided frequency selective
communication systems with hybrid architectures. Different from the
existing deep learning aided network, we combine the data driven network
and the model driven network to jointly enhance spatial and frequency
properties and estimate channels. Meanwhile, we demonstrate how to
leverage attention mechanisms and mobile networks for effective estimating
hybrid IRS aided systems. Simulation results indicate that the proposed
algorithms are capable of attaining significant performance improvements
in terms of accuracy and pilot overhead.

\bibliographystyle{IEEEtran}
\bibliography{myref}

\end{document}